\documentclass[aps,pra,twocolumn]{revtex4}
\usepackage{graphicx}
\usepackage{amsmath,amssymb}
\begin{document}
\title{Probing the sign of on-site Hubbard interaction by 
two-particle quantum walks}
\author{Andrea Beggi, Luca Razzoli, Paolo Bordone}\email{paolo.bordone@unimore.it}
\affiliation{Dipartimento di Scienze Fisiche, Informatiche e Matematiche, Universit\`a 
di Modena e Reggio Emilia, I-41125, Modena, Italy}
\affiliation{Centro S3, CNR-Istituto di Nanoscienze, I-41125, Modena, Italy}
\author{Matteo G. A. Paris}\email{matteo.paris@fisica.unimi.it}
\affiliation{Quantum Technology Lab,
Dipartimento di Fisica dell'Universit\`{a} degli Studi di
Milano, I-20133 Milano, Italy}
\begin{abstract}
We consider two identical bosons propagating on a one-dimensional 
lattice and address the problem of discriminating whether their mutual 
on-site interaction is attractive or repulsive. We suggest a probing 
scheme based on the properties of the corresponding two-particle quantum walks, 
and show that the sign of the interaction introduces specific and detectable 
features in the dynamics of quantum correlations, thus permitting to discriminate 
between the two  cases. We also discuss how these features are connected 
to the band-structure of the Hubbard Hamiltonian, and prove that discrimination
may be obtained only when the two walkers are initially prepared in a superposition 
of localized  states.
\end{abstract}
\date{\today}
\maketitle
\section{Introduction}
The Hubbard model (HM) \cite{essler,montorsi1992hubbard,lieb2003one}
describes the physics of several systems of correlated bosons (or fermions) 
\cite{Amico2008} on different platforms, e.g. ultracold atomic lattices \cite{Preiss2015,Duan2003,fukuhara2013,Jaksch200552},
spin chains \cite{fukuhara2013b,Schulz1990} and nonlinear waveguides
\cite{Lahini2012,Bromberg2010,Lederer2008,Lee2014,longhi2011optical}.
The same model \cite{matsubara1956lattice,fukuhara2013b} may be also 
employed to describe the propagation of  identical (and interacting) 
particles \cite{Zong-Qin2014,wang2014anyons,Wang2015spinflipQW}
on a one-dimensional lattice. In these systems, the on-site interaction 
can be either attractive or repulsive, depending on the chosen physical 
implementation. Remarkably, there are systems where both possibilities 
are contemplated  
\cite{Denschlag2007,Bromberg2010,winkler2006repulsively,wang2010pairquench}.
\par
Recent experimental results with cold atoms  have shown
that a bounded pair of interacting particles may be formed 
even under the action of a repulsive potential\cite{winkler2006repulsively,Preiss2015,folling2007direct,Piil2007,wang2008quantum}.
This phenomenon may be understood by looking at the
band-structure of the model, which behaves symmetrically under the
exchange of sign in the interaction term: in both cases the onset of 
the interaction creates a separate {\em mini-band} hosting bounded states, 
thus determining the co-propagation  of the particles that are initially 
placed on the same site \cite{valiente2008two,Lahini2012}. This feature 
have suggested the hypothesis that the model is fully symmetrical under 
the exchange of sign of  the on-site interaction, and this conjecture was 
somehow corroborated by the fact that no differences may be seen in the 
correlations among particles \cite{Lahini2012}, at least when localized 
initial states are considered.
\par
At variance with these results, more recent studies on the Hubbard dynamics 
of identical particles \cite{schneider2012fermionic,Lee2014,Beggi2016colonoise} 
have shown that depending on the nature of the initial states, the HM may 
indeed lead to different dynamics if we switch the sign of the on-site interaction. 
Besides the fundamental interest, this fact may be useful to engineer 
entanglement at a deeper level, thus granting a further degree of freedom 
to perform (quantum) computational tasks.
\par
Following the considerations above, here we address the discrimination 
between attractive and repulsive on-site interaction in the Hubbard model. 
In particular, we analyze the dynamics of two identical bosons propagating 
on a one-dimensional lattice and show that the sign of the interaction 
clearly influences the evolution of the system, as well as the nature 
of the particle correlations, thus permitting the discrimination between 
the two cases. We also devote attention to prove that i) these features 
are intimately  connected to the band-structure of the Hubbard Hamiltonian 
and ii) discrimination is possible when the two particles are initially 
prepared in a {\em superposition} of localized states, whereas for localised
states no differences may be observed \cite{valiente2008two,Lahini2012}.
\par
Our probing scheme is based on the fact that the Hubbard Hamiltonian
describes the quantum walks of identical particles 
on a one-dimensional lattice. In turn, we consider the two identical 
bosons as quantum probes \cite{qpsp14,frac14,fprobes,siloi17} and the dynamical
feature of their quantum walks as a tool to reveal the sign of the 
interaction. Besides, we will also develop an intuitive picture for their 
behaviour, adding on previous observations \cite{schneider2012fermionic}.
\par
Quantum walks (QWs) describe the dynamics of one or more quantum particles
on a lattice \cite{Kempe2003_QW,Venegas2012_QW}. They show characteristic
quantum features when compared to their classical counterparts, e.g.
ballistic propagation, coherent superposition and interference.
These effects make QWs suitable for the implementation of quantum algorithms \cite{Ambainis2003,portugal2013quantum,venegas2008quantum,dt14}. 
In turn, those potential applications inspired
a series of experiments, especially with 
optical networks \cite{Bromberg2009,Peruzzo2010,Rai2008,Tamascelli2016}, 
see \cite{wang2013physical} for a more comprehensive review.
In particular, experimental realizations of photonic quantum walks 
\cite{Broome2013,Spring2013,Tillmann2013,Peruzzo2010},
have provided suitable architectures that outperform 
their classical counterparts for some specific tasks.
In these systems, QWs are typically performed by identical particles
and indistinguishability of the walkers may, in turn, build up genuinely
quantum correlations even in the absence of interaction between the
particles \cite{Franson2013,Benedetti2012_QWferbos}, as it was observed
experimentally in photonic waveguides \cite{Bromberg2009,Peruzzo2010,Mayer2011}.
\par
In order to characterize the sign of the on-site interaction we analyze
in details the dynamics of the two walkers and the time evolution of their
quantum correlations with focus on entanglement. In turn, despite 
entanglement among identical particles has raised much interest 
\cite{Schliemann2001,Eckert2002,Buscemi2006,Buscemi2007,Zanardi2002,Ghirardi2002,Ghirardi2004,Benatti2014,Wiseman2003,Dowling2006,Sasaki2011,Iemini2013,Iemini_PhysRevA.87.022327,Iemini_PhysRevA.89.032324,Reusch_PhysRevA.91.042324,Zanardi2004,Barnum2004,Benatti2012,Benatti2014b},
no universally accepted measure for its quantification is present.
On the other hand, among the different criteria, the so-called 
\emph{entanglement of particles} \cite{Wiseman2003} is perhaps 
the most convenient, due to its simple computability, and to the 
fact that it may be accessible in practical scenarios \cite{Dowling2006}. 
Indeed, it has been recently employed to quantify quantum correlations in spin
chains \cite{Mazza_1367-2630-17-1-013015,Iemini_PhysRevB.92.075423}, 
where it also detects quantum phase transitions, and in model
systems of QWs described by the Hubbard Hamiltonian \cite{Benedetti2012_QWferbos,Buscemi2011_tripbosfer,Beggi2016colonoise}. 
\par
The paper is organized as follows: in Section \ref{sec:Theoretical-Model}
we introduce the bosonic Hubbard-model, whereas in Section \ref{qc} we briefly 
review  the theoretical tools to quantify entanglement between the walkers. 
In Section \ref{sec:Quantum-Walk-Simulation} we illustrate the behaviour 
of the two walkers by numerically solving the dynamics
for a chain with $N=30$ sites, whereas 
in Section \ref{sec:Discussion} we discuss the results 
with the help of a simplified 
semi-analytical model (a reduced chain with $N=4$ sites). 
Finally, Section \ref{sec:Conclusions} closes the paper with some 
concluding remarks, and Appendix A presents some further discussions 
about the symmetries of the system, in order to better appreciate
the results presented in the body of the paper.
\section{The interaction model and the band-structure\label{sec:Theoretical-Model}}
The Bose-Hubbard Hamiltonian describes a collection of spinless bosons
propagating on a one-dimensional made of $N$ sites, with periodic
boundary conditions ($N+1=1$). The Hamiltonian ($\hbar=1$), it is given by:
\begin{align}
{H}_{N}(J,V)&=-J {h}_{N}(v) \\
{h}_{N}(v)&=-\sum_{i=1}^{N} {c}_{i+1}^{\dagger} {c}_{i}+ {c}_{i}^{\dagger} {c}_{i+1}+\frac{v}{2}\sum_{i=1}^{N} {n}_{i}( {n}_{i}-1) \notag
\label{eq:Bose_hubb_Hamilt}
\end{align}
where $ {c}_{i}$ ($ {c}_{i}^{\dagger}$) is the operator that
destroys (creates) a particle on site $i$ of the chain, $J$ is the
hopping amplitude, $V$ is the strength of the interaction among
the bosons sharing the same site (attractive for $V<0$, repulsive
for $V>0$) and $v=V/J$ is the relative strength of the interaction 
with respect to the hopping energy. In the following we focus on 
a system with total number of particles $n=2$. 
Since $J$ is only a time-dilation factor, giving the
characteristic time of the hopping dynamics, we introduce 
the dimensionless time $\tau=|J|\,t$. The quantity $v$ is thus 
the sole parameter influencing the physics of the system.
\par
The states of the system can be equally represented in the Fock
space and in the symmetrized two-particle Hilbert space, with dimension
$N(N+1)/2$ and basis set given by $\{\left|1,1\right\rangle _{s},\left|1,2\right\rangle _{s},...,\left|2,2\right\rangle _{s},...,\left|N,N\right\rangle _{s}\}$,
where $\left|i,j\right\rangle _{s}$ ($j\ge i$) stands for a symmetrized
state in which one particle is localized on site $i$, and the other
on site $j$ \cite{dt06}. 
Since the Hubbard Hamiltonian represents the discrete
version of the kinetic operator plus a central potential (depending
only on the relative distance among the particles), the dynamics can
be factorized in the coordinate space of the center of mass $R=\frac{i+j}{2}$
and the relative distance $r=j-i$, so the ansatz for its eigenfunctions
becomes
\begin{equation}
\Phi(R,r)=e^{iKR}\varphi(r),\label{eq:EigHamiltBH}
\end{equation}
and the energy spectrum $E(K)=\omega(K)$ depends on the quasimomentum
$K$, which assumes only discrete values due to the periodic boundary
conditions ($K=\frac{2\pi}{N}\nu$, where $\nu=1,\,2,...,\, N$). 
The bandstructure \cite{scott1994quantum} is composed by a \emph{mini-band}, 
hosting $N$ states with energies near $V$, and a \emph{main subband}, extendend
approximately between $-4J$ and $+4J$ and hosting the remaining
$N(N-1)/2$ states\cite{Lahini2012,valiente2008two}. The eigenstates in 
the mini-band are associated with the so-called
\emph{bound states}, in which the particles share the same site and
show a co-walking dynamics. Conversely, the eigenstates of the main
subband, the so-called \emph{scattering states}, have a delocalized
wavefunction (with $\varphi(0)\sim0$) and show a fermionic anti-bunching
behaviour in the high $V$ regime \cite{Lahini2012}. The striking
aspect is that also a repulsive potential may create a bound pair
of bosons, as it has been shown experimentally \cite{winkler2006repulsively}.
Indeed, the evolution of particles which are initialized in a bound
state is dominated by the states of the miniband, which in the high
$V$ regime are well separated in energy from the states of the main
subband: therefore, the bound-state particles are forced to remain
on the same site while performing their quantum walk.
\par
\begin{figure}[h]
\includegraphics[width=0.49\columnwidth]{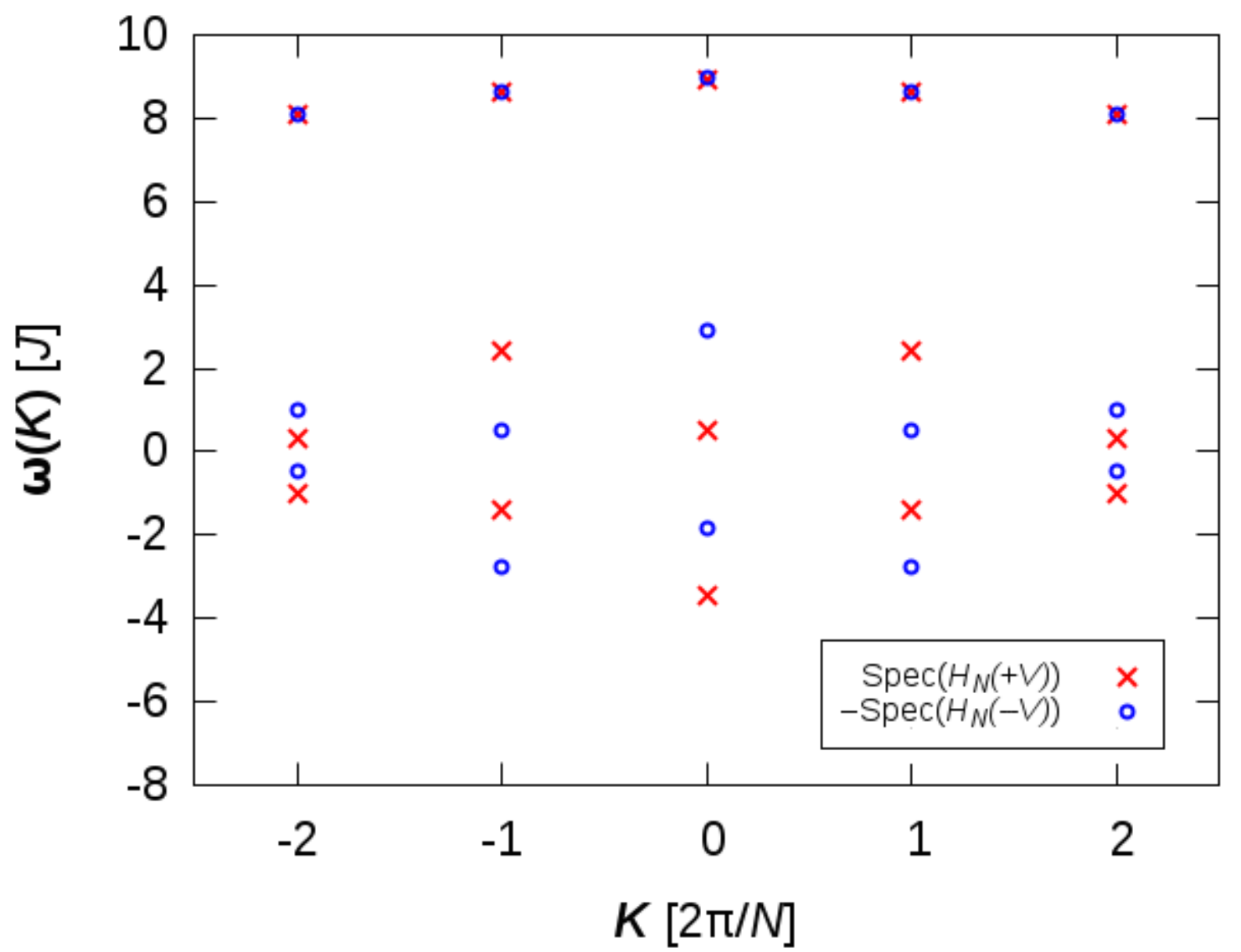}
\includegraphics[width=0.49\columnwidth]{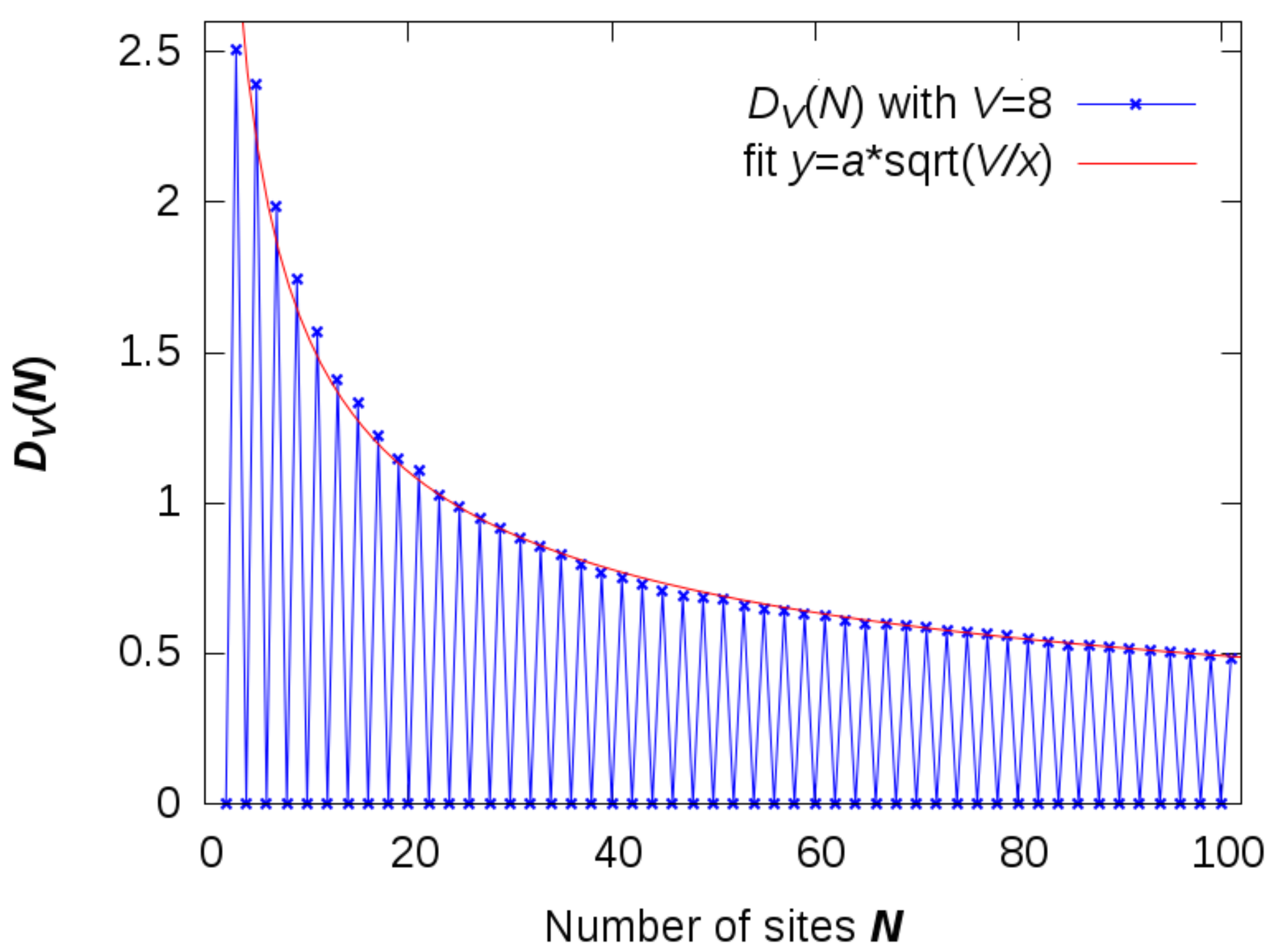}
\caption{(Left): band-structure (spectrum) of $ { {H}}_{N}(V)$ and 
$- { {H}}_{N}(-V)$
for $V=8$ and $N=5$. Energies are given in units of $J$, wave-vectors
in units of $2\pi/N$. (Right): Deviation $D_{V}(N)$ between the 
spectra of $ { {H}}_{N}(V)$
and $ { {H}}_{N}(-V)$ for $V=8$ at different values of
$N$. The two spectra are exactly opposite ($D_{V}(N)=0$) only for
even $N$. The red line is a phenomenological fit for odd $N$ values. }
\label{fig:BandaNDisp}
\end{figure}

It has been observed \cite{Lahini2012,valiente2008two,Valiente2010twobody}
that the energy spectrum $\mathrm{Spec}[ {H}_{N}(J,V)]$ simply
changes its sign when we change the sign of $V$, i.e. $\mathrm{Spec}[ {H}_{N}(J,V)]=-\mathrm{Spec}[ {H}_{N}(J,-V)]$
(we consider inverse ordering for the two spectra) and for this reason,
it was suggested that the dynamics of the system are the same irrespective
of the sign of $V$ \cite{Lahini2012,winkler2006repulsively}. However, 
this symmetry holds only for chains with even
$N$, whereas for chains with odd $N$, as it is shown
in the left panel of Fig. \ref{fig:BandaNDisp} for $N=5$,  the 
two spectra display some clear discrepancies (mostly in the main subband). 
If we refer to the two spectra as $\mathrm{Spec}[ {H}_{N}(J,\pm V)]
=\{\omega_{i}^{\pm}\}_{i}$, we may define the deviation $D_{V}$ between them as: 
\begin{align}
D_{V}(N)& =\left\Vert \mathrm{Spec}[ {H}_{N}(V)]+\mathrm{Spec}[ {H}_{N}(-V)]\right\Vert  \notag \\ 
&=\sqrt{\sum_{i}\left(\omega_{i}^{+}+\omega_{i}^{-}\right)^{2}}.
\end{align}
The behavior of $D_{V}$ against $N$ is reported in the right panel of 
Fig. \ref{fig:BandaNDisp}: $D_{V}$ is always zero for even $N$, while for
odd $N$ it vanishes only in the limit for $N\rightarrow\infty$. This
suggests that the considered effect is related to differences in periodic
boundary conditions, which are less important as soon as $N$ grows. 
Our observations on the effects of $\mathrm{sgn}(V)$ over dynamics
do not depend on $N$ being even or odd. However, in order to avoid
the influence of the asymmetry of the bandstructure, we considered
in our simulations only chains with even $N$. Even in this condition,
however, it will be apparent that there are differences in the dynamics
when switching from attractive to repulsive interactions (or vice
versa).
\section{Two-site correlations and entanglement of particles\label{qc}}
The particle density at each site $i$ of the lattice 
and  the two-particle correlation among sites $i$ and $j$
at a given time $t$ are given by
\begin{align}
n_{i}(t) &  =\left\langle  {c}_{i}^{\dagger} {c}_{i}\right\rangle\,, \\
\Gamma_{i,j}(t) & =\left\langle  {c}_{i}^{\dagger} {c}_{j}^{\dagger} 
{c}_{j} {c}_{i}\right
\rangle \,.
\end{align}
$\Gamma_{i,j}$ coincides with the diagonal element
of the density operator $\rho_{i,j;i,j}={}_{s}\!\left\langle i,j\right| 
{\rho}\left|i,j\right\rangle _{s}$ whereas for $n_{i}$ we have 
$n_i=\sum_{i}\rho_{i,j;i,j}(1+\delta_{i,j})$.
The two-particle correlation is useful to identify behaviours like bunching
(or co-walking) and anti-bunching, which depend both on the initial
state $\left|\Psi(0)\right\rangle $ and on the strength of the interaction
$V$ \cite{Zong-Qin2014,Lahini2012}. 
Here, we will show that 
it depends on $\mathrm{sgn}(V)$ for some initial states. 
In particular, for the sake of 
simplicity we consider the normalized two-particle correlations: 
\begin{equation}
\tilde{\Gamma}_{i,j}(t)=\frac{\Gamma_{i,j}(t)}{\underset{\{i,j\}}
{\max}[\Gamma_{i,j}(t)]}.
\end{equation}
In order to quantify the entanglement among the walkers,
we employ the \emph{entanglement of particles} $E_{P}$
\cite{Wiseman2003}, i.e. 
\begin{equation}
E_{P}=\sum_{k=0}^{n}P_{k,n-k}\, {E}(\rho_{k,n-k})\,,
\label{eq:Enta_parts}
\end{equation}
where, given a bipartition of the sites $A=\{n_{Ai}\}_{i}$ 
and $B=\{n_{Bi}\}_{i}$
\begin{equation}
\rho_{k,n-k}=\Pi_{k,n-k}\rho\Pi_{k,n-k}
\end{equation}
is the projection of the system state $\rho$ over the subspace in
which $A$ contains exactly $k$ particles and $B$ the remaining
$n-k$ ones, whereas $P_{k,n-k}=\mathrm{Tr}[\rho_{k,n-k}]$ is
the corresponding probability. For any given 
bipartition the projectors $\Pi_{k,n-k}$ may be expressed as 
\begin{align}
\Pi_{k,n-k} 
 & = \!\!\!\sum_{\Sigma_{i}n_{Ai}=k}\!\!\!\!
 \left|\{n_{Ai}\}\right\rangle \left\langle \{n{}_{Ai}\}
 \right|\otimes\!\!\!\!\!
 \sum_{\Sigma_{i}n_{Bi}=n-k}\!\!\!\!\!\!
 \left|\{n_{Bi}\}\right\rangle \left\langle \{n{}_{Bi}\}\right|\,,\nonumber 
\end{align}
and satisfy the completeness relation gives $\sum_{k=0}^{n}\Pi_{k,n-k}= {\mathbb I}$.
The quantity $ {E}$ can be any standard measure of bipartite
entanglement among the registers individuated by the two partitions. 
For a two-particle system, the terms with $k=0$ or
$k=2$ give zero correlations $ {E}=0$ and thus 
Eq. (\ref{eq:Enta_parts}) reduces to
\begin{equation}
E_{P}=P_{1,1}\, {E}(\rho_{1,1}).
\end{equation}
Since we are considering a Hamiltonian system, the states remain
pure during all their evolution, and we can thus use for  
the \emph{linear entropy}\cite{Buscemi2007} for $ {E}$
\begin{equation}
 {E}(\rho_{AB})=\frac{N}{N-2}\left(1-\mathrm{Tr}_{A}[\rho_{A}^{2}]\right),
\end{equation}
where $\rho_{A}=\mathrm{Tr}_{B}[\rho_{AB}]$
is the reduced density matrix of the subsystem $A$. Equivalent
results may be obtained upon employing the von
Neumann entropy  or the negativity
\cite{Girolami2011negat,Lee2003negat}.
\par
It is worth noting that $E_{P}$ depends upon the chosen bipartition
of the system modes among Alice and Bob: therefore, different partitions
of the system can lead to different values of $E_{P}$. Also, it should
be mentioned that $E_{P}$ does not capture all the quantum correlations
encoded in the system: indeed, ``ideal'' co-walking situations,
where the particles are strongly correlated, do not give contributions
to $E_{P}$ (since they correspond to $k=0$ or $k=2$). On the other hand,
those states cannot be exploited to perform any task,
since one of the two observers is left with no particle on which she can 
perform any local operation. $E_P$ thus appears to capture the presence 
of quantum correlations that represent a resource for quantum information
processing.
\section{Probing the sign by two-particle quantum walks
\label{sec:Quantum-Walk-Simulation}}
\begin{figure}[h!]
\includegraphics[width=0.9\columnwidth]{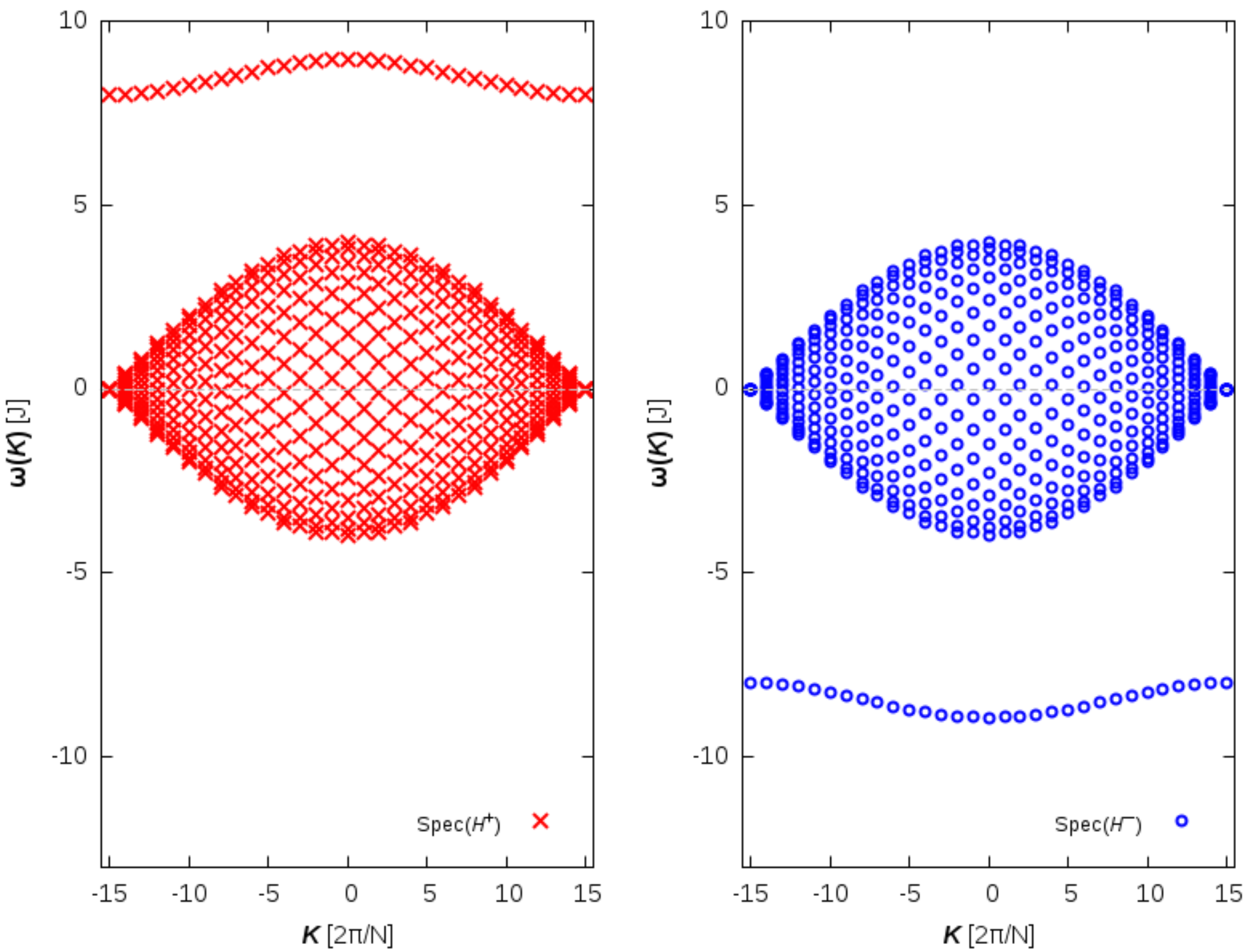}
\caption{Bandstructure $\omega(K)$ of a chain with $N=30$ sites at $V/J=+8$
(left, red) and $V/J=-8$ (right, blue). }
\label{wrap:fig:BandaN30}
\end{figure}

In this section we describe how the sign of the interaction may be revealed by
the features of the walkers' dynamics and by their correlations. 
As a representative situation, we choose a lattice with $N=30$ sites. 
The dynamics of the system is driven either
by the Hamiltonian $ { {H}}_{+}= { {H}}_{N}(J,V)$
or $ { {H}}_{-}= { {H}}_{N}(J,-V)$, where $J=1$. 
Their spectra $\omega(K)$ (which are symmetrical with
respect to $\omega=0$) are reported in Fig. \ref{wrap:fig:BandaN30}.
In studying the dynamics, we limit ourselves to interaction times 
such that the two particles remain far from the boundaries
of the lattice, in order to avoid interference effects due to periodic
boundary conditions.
Given the initial preparation $ {\rho}(0)=\left|\Psi(0)\right\rangle 
\left\langle \Psi(0)\right|$ of the two particles we evaluate the evolved
state ${\rho}(t)= { {U}_\pm}(t) {\rho}(0) {
 {U}_\pm}^{\dagger}(t)$,
$ { {U}_\pm}(t)=\exp(-i { {H}}_{_\pm}t)$ by numerical 
diagonalization of the Hamiltonian(s) \cite{gpu17}.
\par
As possible initial states $\left|\Psi(0)\right\rangle $ we consider 
the following ones:
\begin{eqnarray}
\left|\Psi_{1}\right\rangle  & = & \left|15,17\right\rangle _{s},\\
\left|\Psi_{2}\right\rangle  & = & \left|14,16\right\rangle _{s},\\
\left|\Psi_{3}\right\rangle  & = & \left|14,17\right\rangle _{s},\\
\left|\Psi_{4}\right\rangle  & = & \left|14,16\right\rangle _{s}+\left|15,17\right\rangle _{s},\\
\left|\Psi_{5}\right\rangle  & = & \left|14,16\right\rangle _{s}+\left|14,17\right\rangle _{s},
\end{eqnarray}
which we evolve, alternatively, under the action of $ { {U}}_{\pm}(t)$.
\par
\begin{figure}[h!]
\includegraphics[width=0.98\columnwidth]{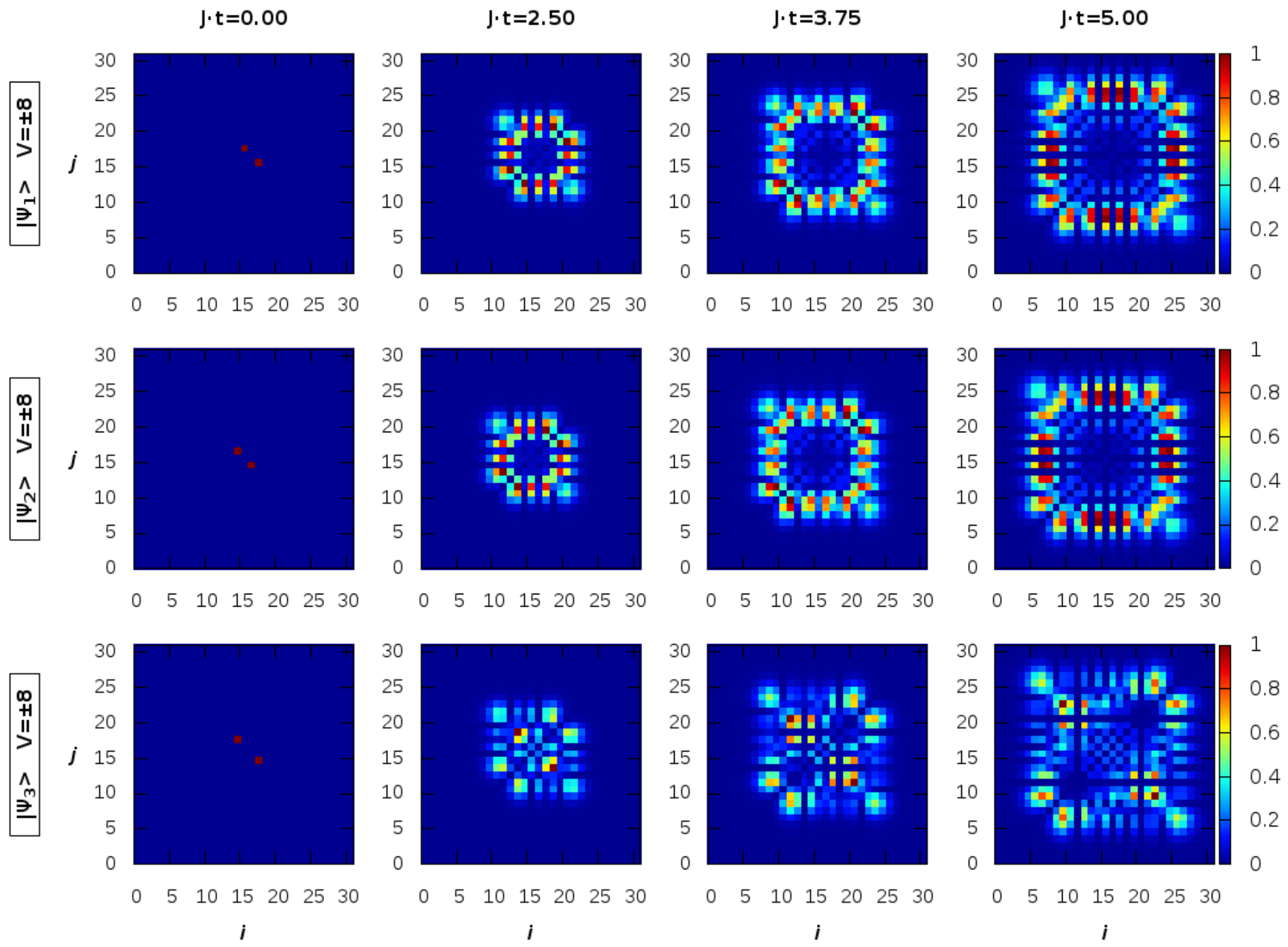}
\includegraphics[width=0.98\columnwidth]{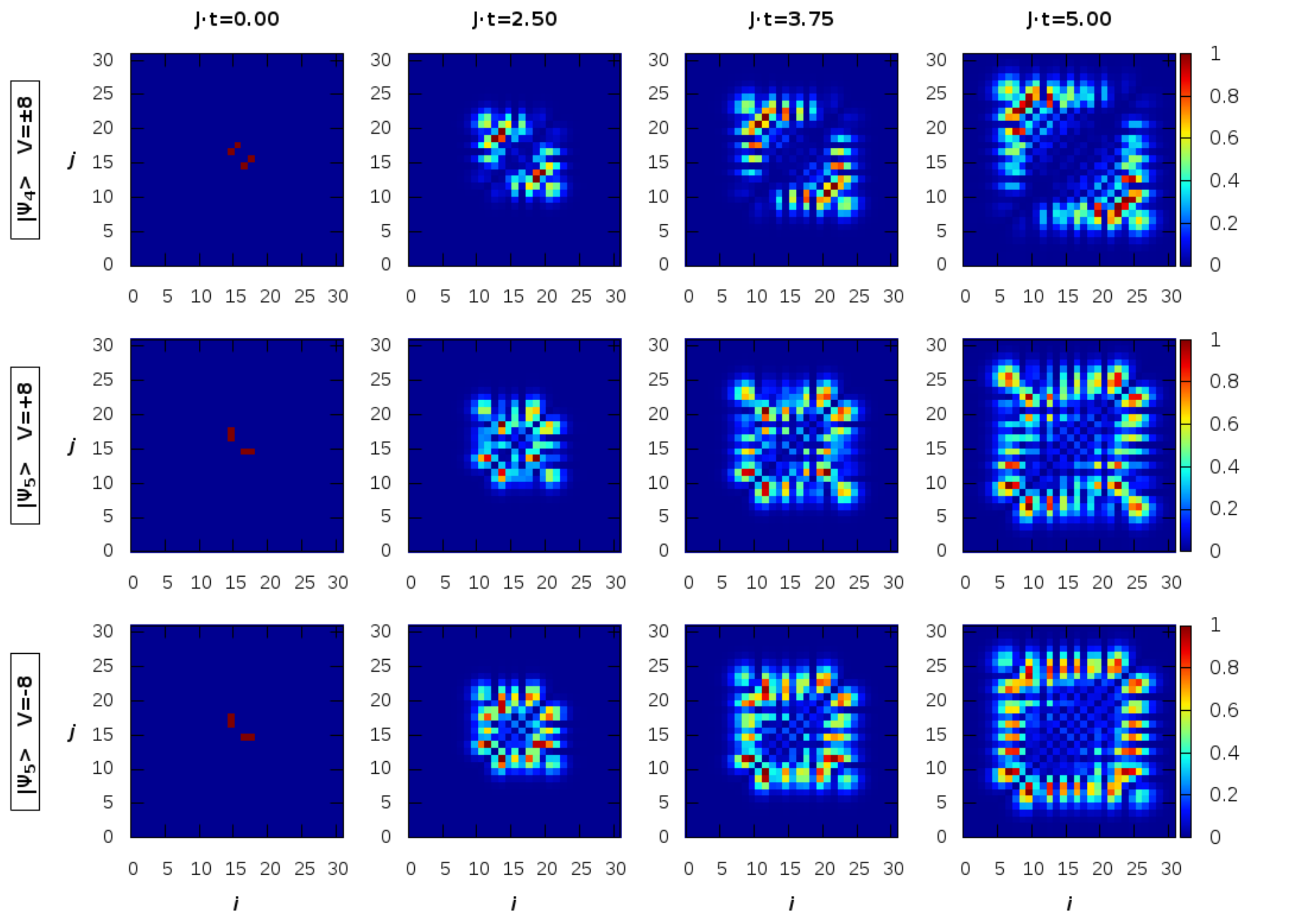}
\caption{(Upper panel): Evolution of two-sites correlations 
$\tilde{\Gamma}_{i,j}$ in states
$\left|\Psi_{1}\right\rangle ,\left|\Psi_{2}\right\rangle $ 
and $\left|\Psi_{3}\right\rangle $
under $ { {H}}_{+}$ and $ { {H}}_{-}$, with
$V=\pm8$ and $J=1$.
(Lower panel): the same for states
$\left|\Psi_{4}\right\rangle $ and $\left|\Psi_{5}\right\rangle $.
\label{fig:Psi12345}}
\end{figure}
\begin{figure}[h]
\includegraphics[width=0.49\columnwidth]{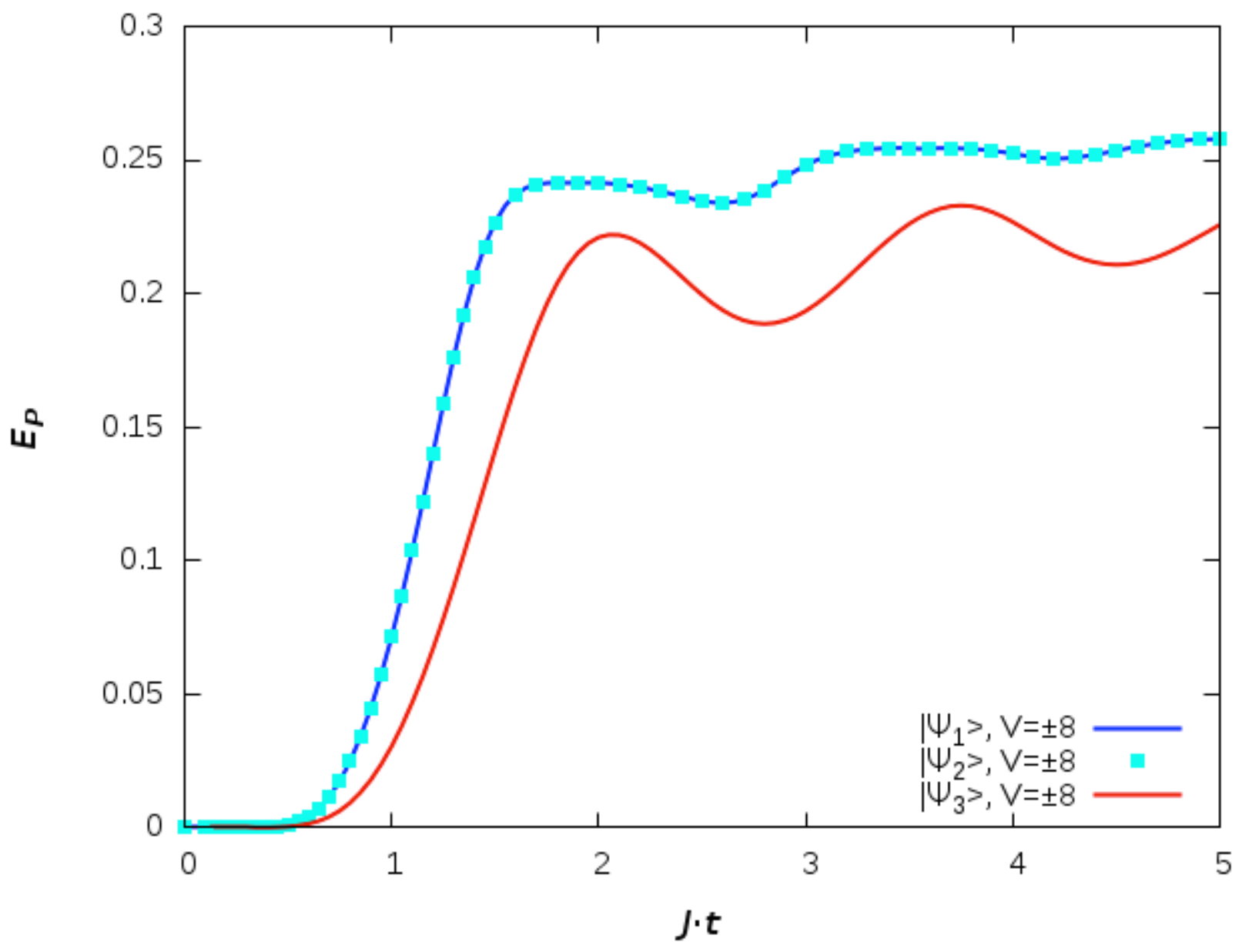}
\includegraphics[width=0.49\columnwidth]{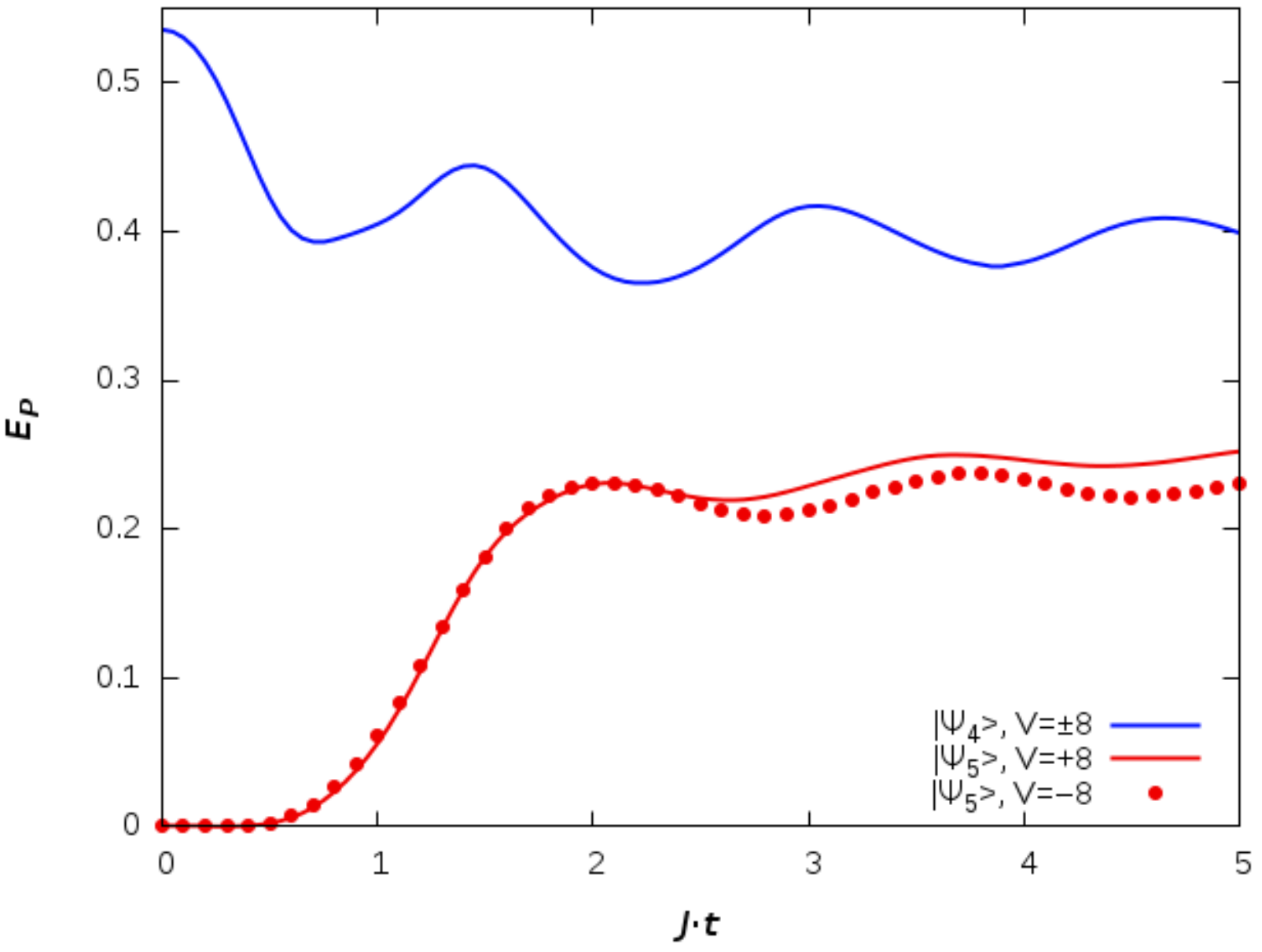}
\includegraphics[width=0.49\columnwidth]{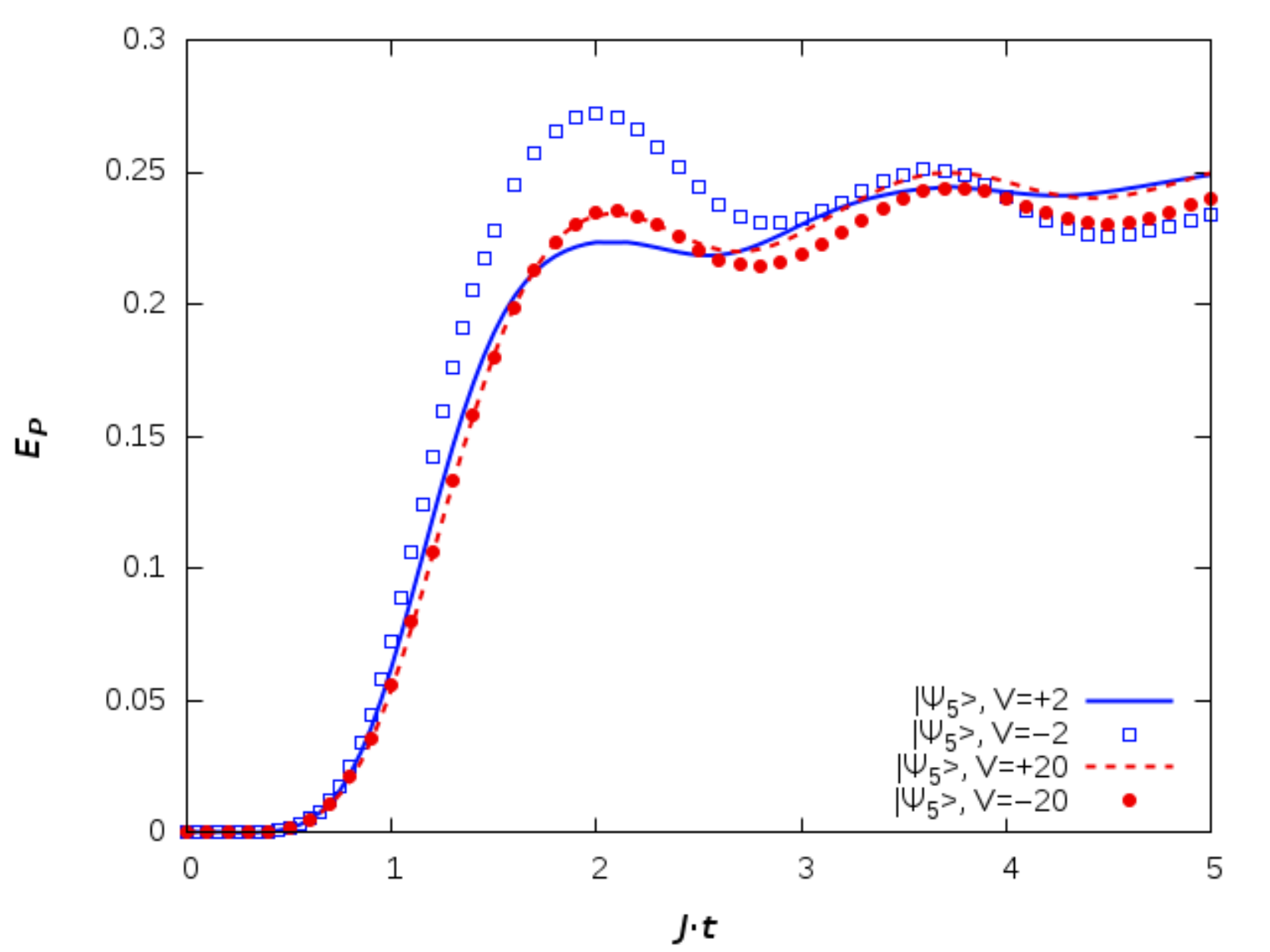}
\includegraphics[width=0.49\columnwidth]{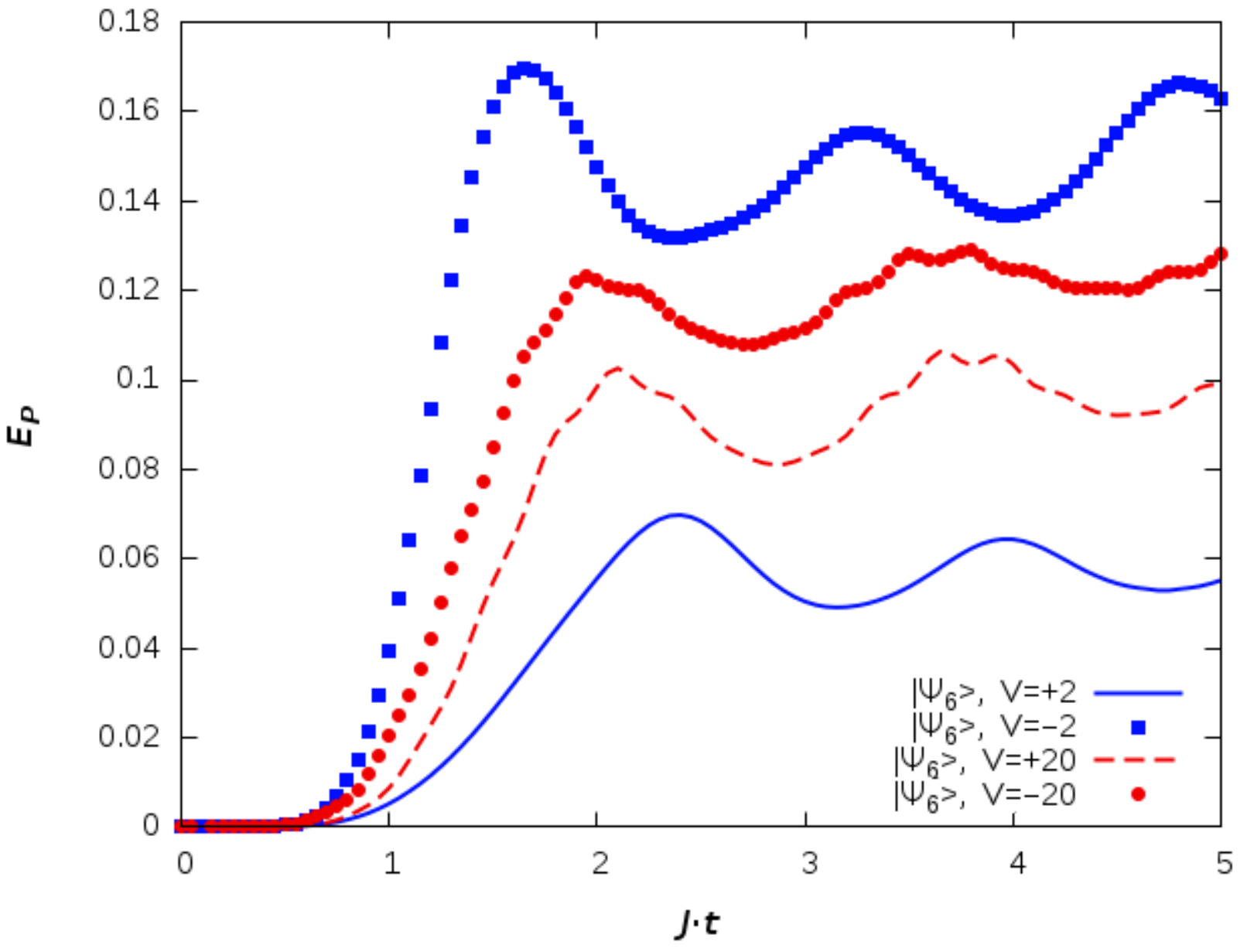}
\caption{Time evolution of entanglement of particles under $ { {H}}_{\pm}$. 
(Upper left):  states $\left|\Psi_{1}\right\rangle ,\left|\Psi_{2}\right\rangle$, $V=\pm 8$. 
(Upper Right): states $\left|\Psi_{5}\right\rangle $ under $ { {H}}_{\pm}$, $V=\pm 8$.
(Lower left): state $\left|\Psi_{5}\right\rangle $ and different values of $V$. 
(Lower right): state $\left|\Psi_{6}\right\rangle $ and different values of $V$.\label{fig:PsiEnta}}
\end{figure}
\begin{figure}[h]
\includegraphics[width=\columnwidth]{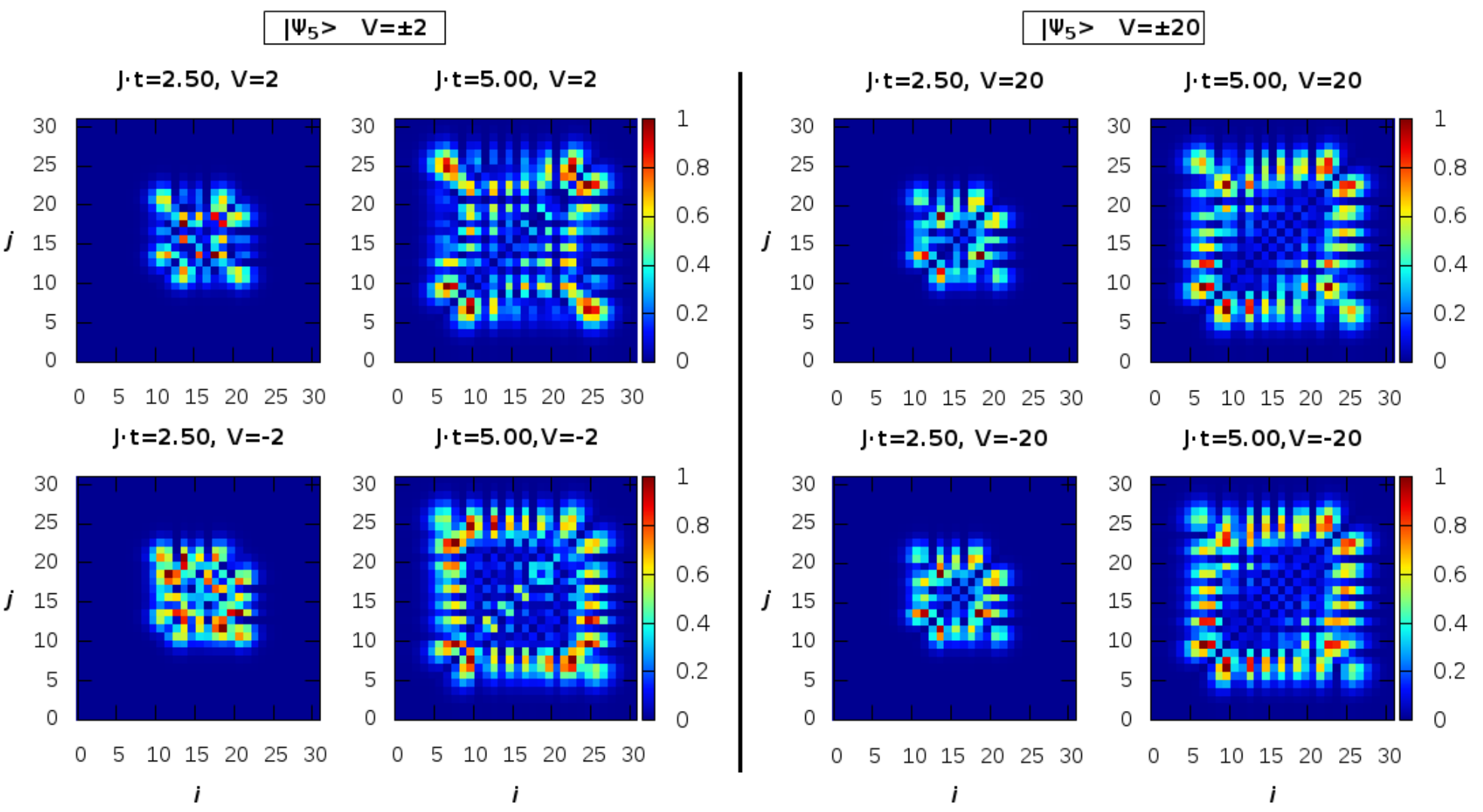}
\caption{Evolution of two-sites correlations $\tilde{\Gamma}_{i,j}$ in state
$\left|\Psi_{5}\right\rangle $ under $ { {H}}_{+}$ and
$ { {H}}_{-}$ for different values of $V$ and for $J=1$.
Differences between evolutions are more apparent at low potential
energies (i.e. $V\sim J$)\protect \\
\protect \\
}
\label{fig:Psi5-2-20}
\end{figure}

Let us start by looking at the behaviour of the correlations 
$\tilde{\Gamma}_{i,j}$. As it can be seen in the upper panel of Fig. 
\ref{fig:Psi12345}, $\tilde{\Gamma}_{i,j}$ for the states 
$\left|\Psi_{1}\right\rangle $, $\left|\Psi_{2}\right\rangle $
and $\left|\Psi_{3}\right\rangle $ is invariant under the sign exchange
of $V$. The evolution corresponds to the choice $V=\pm8$, simulations 
with different values of $V$ lead to the same behavior. Notice that 
the evolutions of $\left|\Psi_{1}\right\rangle $
and $\left|\Psi_{2}\right\rangle $ are practically identical, except
for a rigid shift, and this makes sense since $ { {H}}_{N}$
with periodic boundary conditions commutes with the translation operator
$ {T}_{l}=\Sigma_{i}\left|i+l\right\rangle \left\langle i\right|$,
therefore two states that differ only for a rigid shift of $l=1$
sites ($\left|\Psi_{1}\right\rangle = {T}_{1}\left|\Psi_{2}\right\rangle $)
should have the same dynamics. The same behaviour is found for the entanglement
of particles $E_P$, which is identical both for $ { {H}}_{+}$
and $ { {H}}_{-}$ and is reported in the upper panels of 
Fig. \ref{fig:PsiEnta}; $E_P$ is initially zero since all states are symmetrized
versions of factorizable states. 
Notice that those initial states
are eigenstates of the number operators $ {n}_{i}$, i.e. they
have an exact number of particles in each site of the lattice. The evolution
of these states is invariant when switching from $ { {H}}_{+}$
to $ { {H}}_{-}$ \cite{Lahini2012}. However, the same consideration does
not hold, in general, for superpositions of these states 
\cite{schneider2012fermionic,Lee2014}: as it can be seen in the lower 
panel of Fig. \ref{fig:Psi12345}, the evolution
of correlations $\Gamma_{i,j}$ in $\left|\Psi_{4}\right\rangle $
is the same for $V$ and $-V$, whereas the evolution of correlations
in $\left|\Psi_{5}\right\rangle $ is appreciably different for attractive
and repulsive interactions. 
\\
The effects of the interaction sign may be seen also in the
dynamics of entanglement of particles (see the upper panels of 
Fig. \ref{fig:PsiEnta}): $E_P$ for  $\left|\Psi_{4}\right\rangle $ is
independent from the sign of $V$, while the entanglement of 
$\left|\Psi_{5}\right\rangle $ is different for $V$ and $-V$ 
(notice that $E_{P}(\left|\Psi_{5}\right\rangle )$
is initially 0 since the state is factorizable).
Further analysis show that this effect depends
on the modulus of $|V|$: in Fig. \ref{fig:Psi5-2-20}, we see that 
differences in $\tilde{\Gamma}_{i,j}$ for $\left|\Psi_{5}\right\rangle $ 
are slightly more marked at $V=\pm2$ than at $V=\pm20$, and the same behaviour
is found for entanglement. Indeed,
for some states, e.g.
\begin{equation}
\left|\Psi_{6}\right\rangle =\left|14,14\right\rangle _{s}+\left|14,17\right\rangle _{s},
\end{equation}
at low interaction energy $V/J=\pm2$, $E_P$ may differ by  factor $2$, 
whereas the difference is significantly reduced at higher $V$ 
(see lower panels of Fig. \ref{fig:PsiEnta}).
For this particular state, also the differences in correlations $\tilde{\Gamma}_{i,j}$
(here not reported) are more striking at a lower interaction energy
(they are 10 times larger at $V=\pm2$ compared to what may be seen at $V=\pm20$).
\section{A discussion based on an analytic toy-model with few sites\label{sec:Discussion}}
\begin{figure}[h!]
\includegraphics[width=0.95\columnwidth]{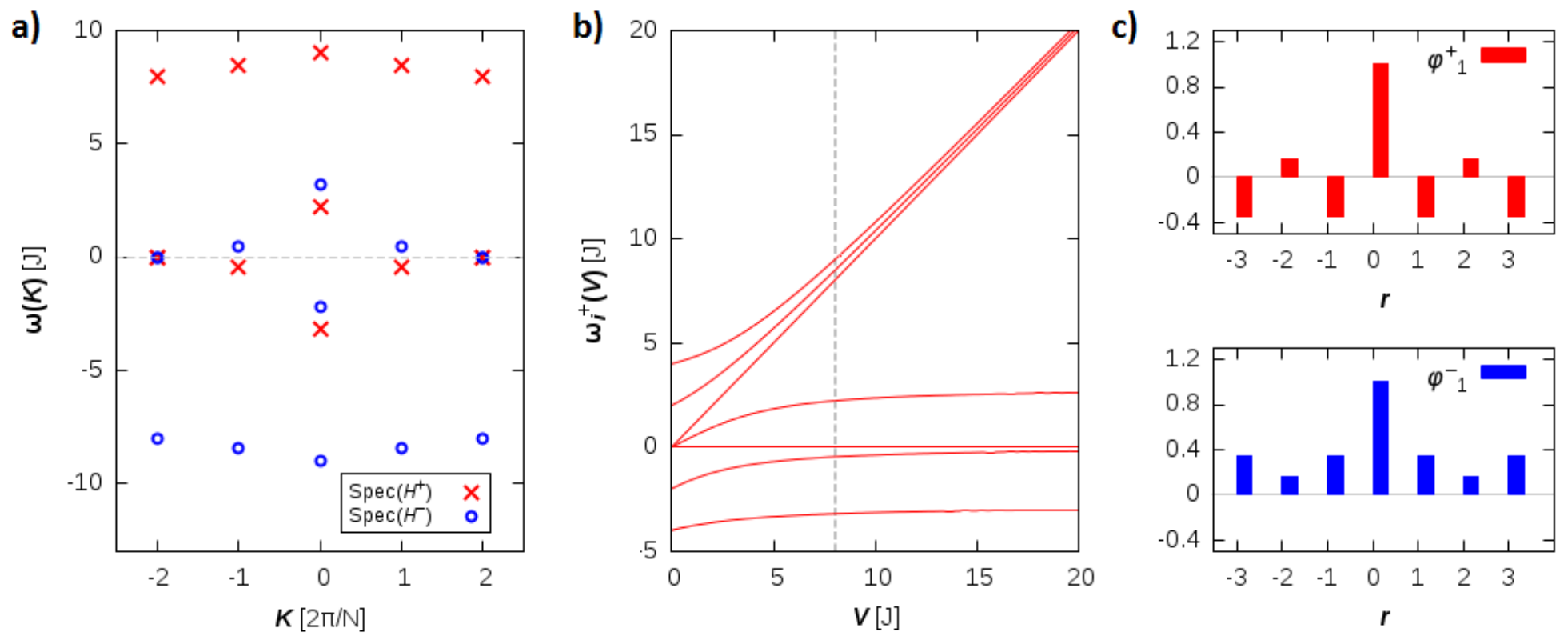}
\caption{\textbf{a)} Bandstructures for $ { {H}}_{+}$ (red crosses)
and $ { {H}}_{-}$ (blue dots) at $V=\pm8$. The main subband
and the miniband are clearly visible, as well as the symmetry of the
two spectra. \textbf{b)} Evolution of the eigenvalues $\omega_{i}^{+}$
at increasing $V$: for high values of the potential energy, there
is a clear separation between the miniband (which becomes almost flat)
and the main subband, while at low $V$ the two subbands are entwined.
The grey vertical dashed line indicates $V=8$ (subfig. a). \textbf{c)}
Radial wavefunction for the first eigenstate of $ { {H}}_{+}$
(upper, red) and $ { {H}}_{-}$(lower, blue).\label{fig:Model4sites}}
\end{figure}

\begin{figure}[h!]
\includegraphics[width=0.99\columnwidth]{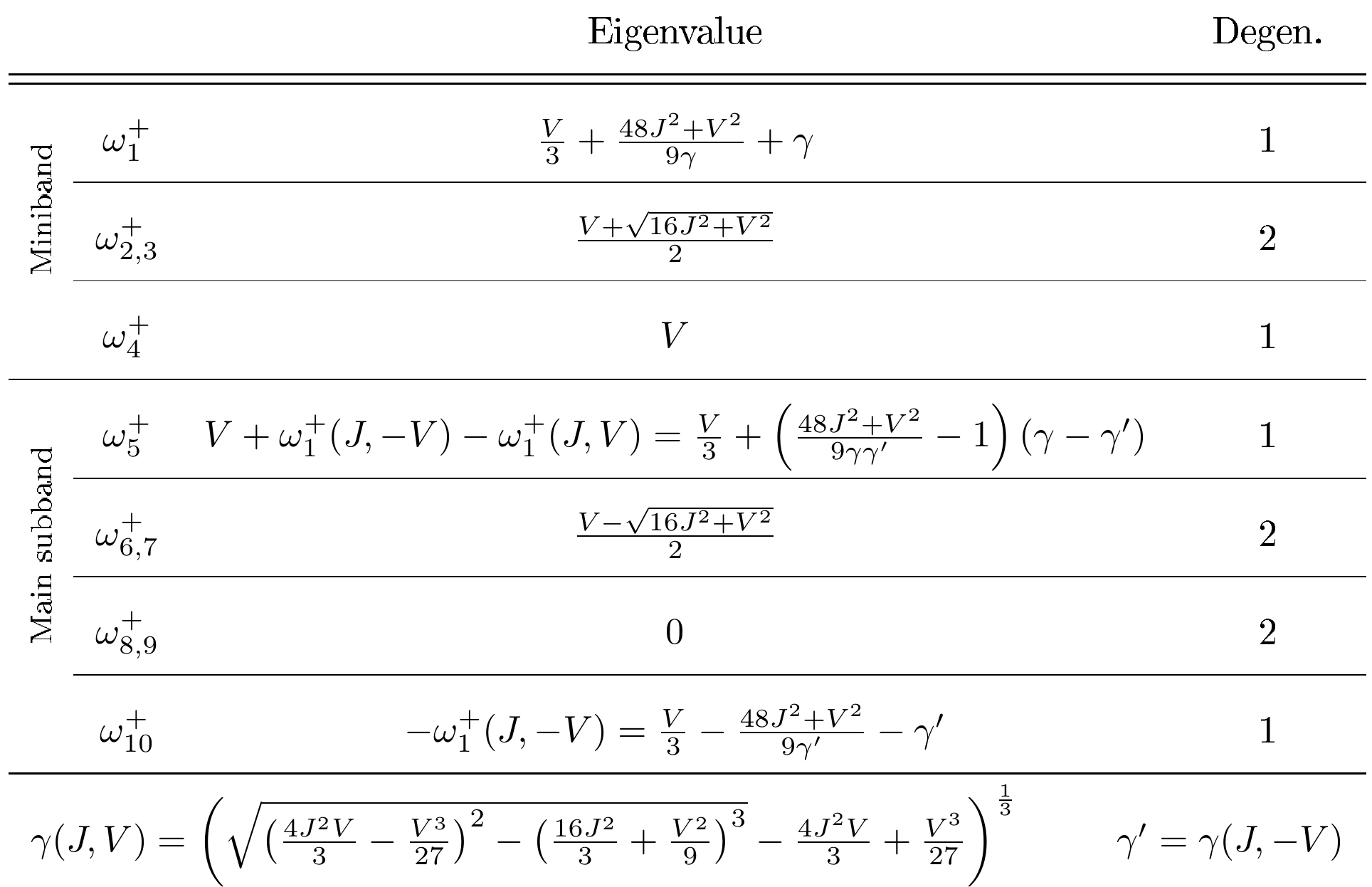}
\caption{Eigenvalues of $ { {H}}_{+}$ with relative multiplicity
(degeneracy), ordered by decreasing energy.}
\label{ftab}
\end{figure}

In order to better understand the behaviour of the system,
and build an intuitive picture,
let us consider an analytic toy-model with a chain 
made of $N=4$ sites. In the left panel of Fig. \ref{fig:Model4sites} we show
the band-structure of $ { {H}}_{\pm}= { {H}}_{4}(J,\pm V)$ 
with $J=1$ and $V=8$, whereas the behaviour of
the eigenvalues of $ { {H}}_{+}$ as a function of $V$ is shown
in the central panel. The eigenvalues of $ { {H}}_{+}$ are denoted
by $\omega_{i}^{+}$ and are reported in Fig. \ref{ftab}, while the eigenvalues
of $ { {H}}_{-}$ can be obtained by replacing
$\omega_{i}^{-}=-\omega_{i}^{+}$
However, since the system is invariant under time-reversal, any change
in the dynamics cannot be related to the simple sign switching of
the eigenvalues.
The radial part of the eigenfunctions, see Eq. (\ref{eq:EigHamiltBH}),
changes sign in some components when switching from $ { {H}}_{+}$
to $ { {H}}_{-}$ \cite{valiente2008two}, while others remain unchanged,
suggesting that the different behaviours with
attractive/repulsive interactions depend on this feature 
of the eigenstates.

For what concerns the number eigenstates, due to the translational
invariance of the chain, the only states which are physically different
are $\left|1,1\right\rangle _{s}$, $\left|1,2\right\rangle _{s}$
and $\left|1,3\right\rangle _{s}$. Each of these states can be decomposed
by projection on the eigenstates of the Hamiltonian, leading to
\begin{equation}
\left|j,k\right\rangle _{s}=\sum_{i}C_{j,k,i}^{\pm}\left|\Phi_{i}^{\pm}\right\rangle .
\end{equation}
 \\
Since the Hamiltonian $ { {H}}_{\pm}$ are real, it is always
possible to choose a set of eigenstates $|\Phi_{i}^{\pm}\rangle$
having real components in the number states basis, therefore the scalar
products 
\begin{equation}
C_{j,k,i}^{\pm}=\left\langle \Phi_{i}^{\pm}\middle|j,k\right\rangle _{s}
\end{equation}
that we obtain with the projection are real numbers too. 
\par
\begin{figure}[h!]
\includegraphics[width=0.98\columnwidth]{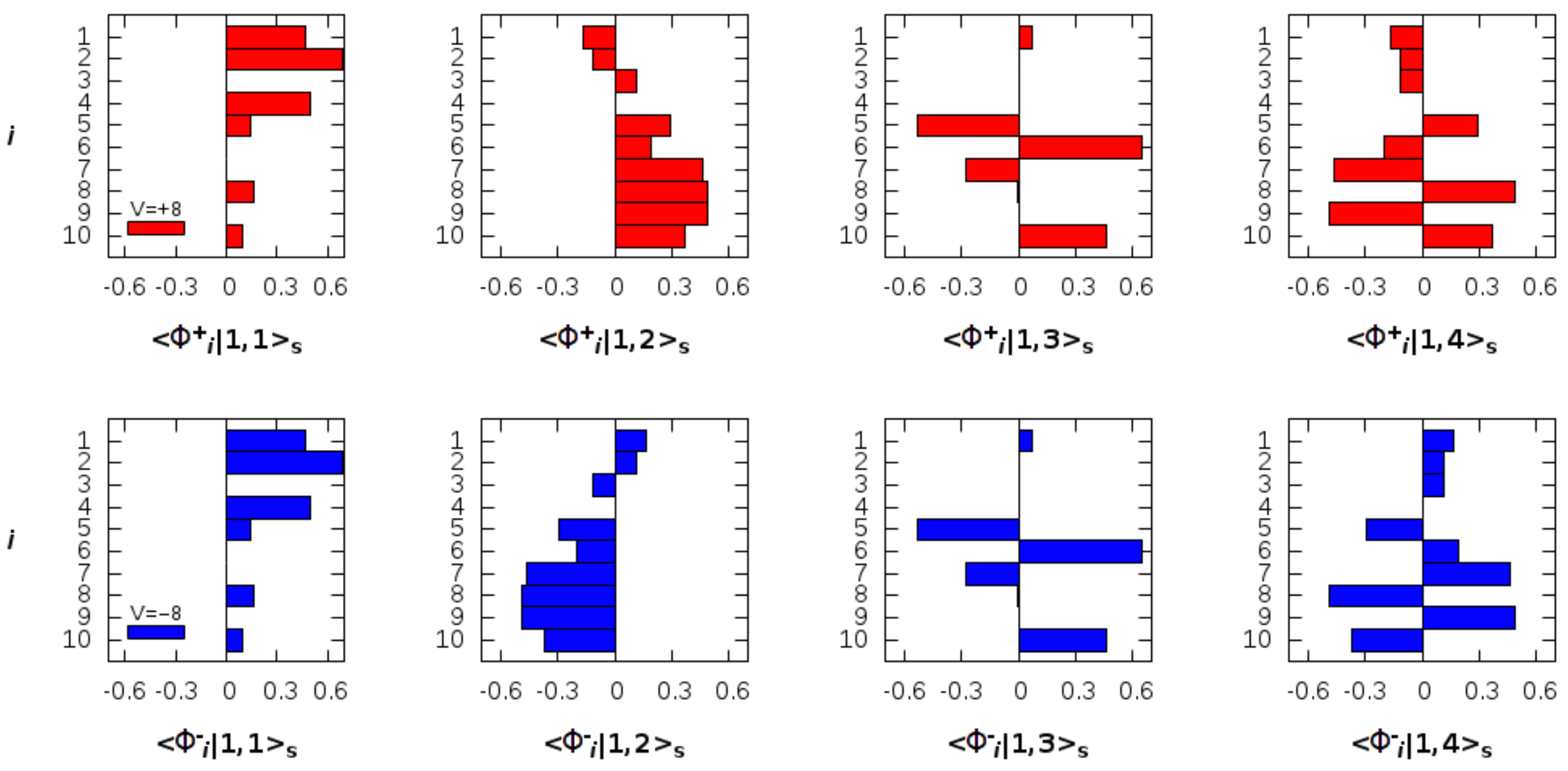}
\caption{Projections of number states $\left|1,1
\right\rangle _{s}$, $\left|1,2\right\rangle _{s}$,
$\left|1,3\right\rangle _{s}$ and $\left|1,4\right\rangle _{s}$
over the eigenstates of $ { {H}}_{+}$ (first row, red)
and $ { {H}}_{-}$ (second row, blue) for a potential energy
of $|V|=8$. Notice that, as expected, $\left|1,4\right\rangle _{s}$
behaves like $\left|1,2\right\rangle _{s}$, and their projections
are also identical in modulus (since the two states are translationally
equivalent).}
\label{fig:ProjV8}
\end{figure}

As you can see in Figure \ref{fig:ProjV8}, all the projections of
states like $\left|1,1\right\rangle _{s}$ and $\left|1,3\right\rangle _{s}$
do not change sign when we switch from $ { {H}}_{+}$ to
$ { {H}}_{-}$, while the projections of states like $\left|1,2\right\rangle _{s}$
do change all their signs. This is the reason why superpositions of
(translationally) equivalent number eigenstates, like
$\left|\Psi_{B}\right\rangle =\left|1,3\right\rangle _{s}+\left|2,4\right\rangle _{s}$,
have the same correlations and entanglement independently from the sign
of $V$, while superpositions of non-equivalent number eigenstates,
like 
$\left|\Xi_{B}\right\rangle =\left|1,4\right\rangle _{s}+\left|2,4\right\rangle _{s}$,
have a different evolution of their correlations under $ { {H}}_{+}$
and $ { {H}}_{-}$, as it was shown in Ref.~\cite{Beggi2016colonoise}.
Indeed, the change of sign in $V$ introduces a relative phase between
the components of the two number eigenstates, which is different for
$ { {H}}_{+}$ and $ { {H}}_{-}$, thus leading
to different evolutions of the states. 
\par
As an example, we have
\begin{align}
 { {U}}(t)&\left(\left|1,3\right\rangle _{s}+\left|2,4\right\rangle _{s}\right)  =  
 \notag \\ &
 \left\{ \begin{array}{ccc}
\sum_{i}\left(C_{1,3,i}^{+}+C_{2,4,i}^{+}\right)|\Phi_{i}^{+}(t)\rangle &  & V>0,\\
\sum_{i}\left(C_{1,3,i}^{+}+C_{2,4,i}^{+}\right)|\Phi_{i}^{-}(t)\rangle &  & V<0,
\end{array}\right.\\
 { {U}}(t)&\left(\left|1,4\right\rangle _{s}+\left|2,4\right\rangle _{s}\right) =  
 \notag \\ & \left\{ \begin{array}{ccc}
\sum_{i}\left(C_{1,4,i}^{+}+C_{2,4,i}^{+}\right)|\Phi_{i}^{+}(t)\rangle &  & V>0,\\
\sum_{i}\left(-C_{1,4,i}^{+}+C_{2,4,i}^{+}\right)|\Phi_{i}^{-}(t)\rangle &  & V<0,
\end{array}\right.
\end{align}
It should be emphasized that the coefficients $C_{j,k,i}^{\pm}$ 
changes sign when we change the sign of $V$, therefore we cannot
observe this effect directly when we study a single number eigenstate
(e.g., by setting $\left|\Psi(0)\right\rangle =\left|1,3\right\rangle _{s}$),
because the squared modulus of the projections do not depend on $\mathrm{sgn}(C_{j,k,i}^{\pm})$.
Only superpositions of number eigenstates keep trace of the sign of
the interaction  \cite{Lee2014,schneider2012fermionic}.
This is perhaps the reason why many previous works in the literature
did not observe this dependence.
Indeed, going back to our calculations on the 1D chain with $N=30$,
we notice that states like $\left|\Psi_{1}\right\rangle $ and $\left|\Psi_{2}\right\rangle $
are of the same kind, i.e. second nearest-neighbours like $\left|i,i+2\right\rangle _{s}$
(this is the reason why they show the same evolution of two-site correlations
- except for a rigid translation), while the state $\left|\Psi_{3}\right\rangle $
is a third nearest-neighbour state, like $\left|i,i+3\right\rangle _{s}$.
This is the reason why $\left|\Psi_{1}\right\rangle $ and $\left|\Psi_{2}\right\rangle $
have equal projections on the eigenstates of $ {H}_{\pm}$$,$
whereas the projections of $\left|\Psi_{3}\right\rangle $ are different.
Therefore, the linear combination of $\left|\Psi_{1}\right\rangle $
and $\left|\Psi_{2}\right\rangle $ is invariant under the exchange
of $\mathrm{sgn}(V)$, while the linear combination of $\left|\Psi_{2}\right\rangle $
and $\left|\Psi_{3}\right\rangle $ is not (some projections exchange
sign for a state but not for the other).
\par
\begin{figure}
\includegraphics[width=0.9\columnwidth]{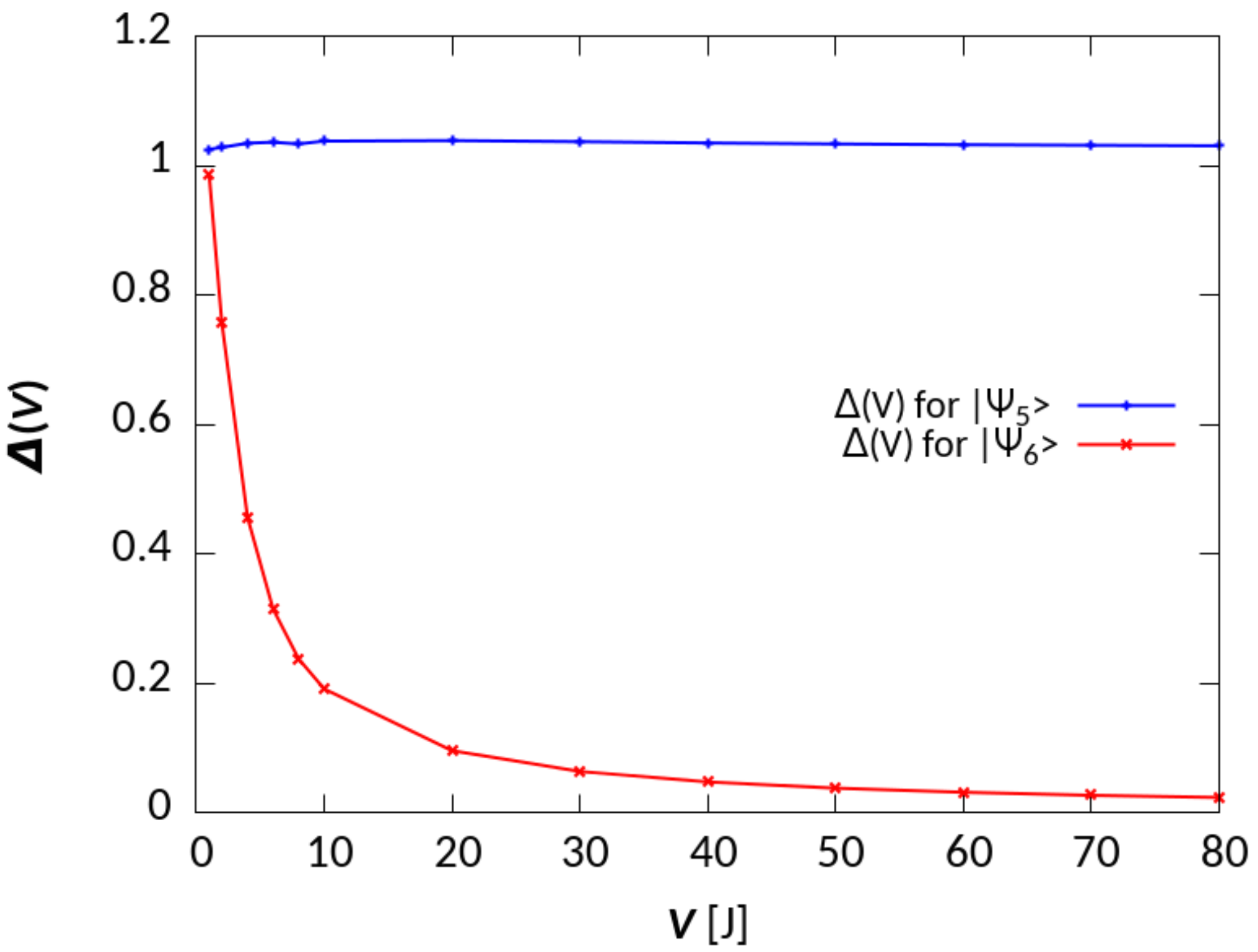}
\caption{Total difference $\Delta(V)$ among the projections of a state $\left|\Psi\right\rangle $
over the eigenstates of $ {H}$ for positive and negative $V$. }
\label{fig:Delta_V_projs}
\end{figure}

Concerning the behaviour of a state like $\left|\Psi_{6}\right\rangle $,
the effect of switching the sign of $V$ is more visible at low potential
energy, since for $V\sim J$ the main subband and the miniband are
strongly mixed, such that each number eigenstate has a projection
over all the eigenvectors of $ { {H}}_{\pm}$, and states
with different nature have therefore very different superpositions
under $ { {H}}_{+}$ and $ { {H}}_{-}$. On the
contrary, for large values of $V/J$ the entanglement is not so much
different, due to the strong separation between the subband
and the miniband (see Fig. \ref{fig:Model4sites}). Indeed, this
separation implies that the projections of the bound state
 $\left|14,14\right\rangle _{s}$
are almost completely contained in the miniband, while the ones of
the scattering state $\left|14,17\right\rangle _{s}$ are mainly in
the main subband. Therefore, the expressions $C_{14,14,i}^{\pm}+C_{14,17,i}^{\pm}$
do not change significantly their (absolute) values when switching
from $ { {H}}_{+}$ to $ { {H}}_{-}$ since, for
each $i$, only one coefficient in the sum is significantly different
from zero. 
 \\
This phenomenon may be better illustrated upon introducing a quantity $\Delta(V)$
that quantifies the difference between the projections of a state $\left|\Psi\right\rangle $
over the eigenstates of $ { {H}}_{+}$ and $ { {H}}_{-}$
as a function of $V$. To this aim we first consider the sum the squared modulus 
of all the projections over the eigenstates with degenerate energies, i.e.
$P^{\pm}(\omega)=\sum_{\omega_{i}=\omega}|\left\langle \Phi_{i}^{\pm}\middle|
\Psi\right\rangle |^{2}$. Then, for each projection $P(\omega)$ on a degenerate 
energy subspace, we evaluate the absolute difference between $P$ at positive 
and negative $V$:
\begin{equation}
\Delta_{\omega}(V)=\left|P^{+}(\omega)-P^{-}(\omega)\right|^{2},
\end{equation}
in order to determine how much the projections of $\left|\Psi\right\rangle $
changes when switching from $\left|\Phi_{i}^{+}\right\rangle $ to
$\left|\Phi_{i}^{-}\right\rangle $. Finally, we sum over the different
energies of the Hamiltonian, and we get the desired figure of merit:
\begin{equation}
\Delta(V)=\sum_{\omega}\Delta_{\omega}(V).
\end{equation}
As we can observe in Fig. \ref{fig:Delta_V_projs}, the quantity $\Delta(V)$
goes to zero for the state $\left|\Psi_{6}\right\rangle $,  
due to the progressive separation between the scattering
subband and the miniband at increasing values of $V/J$. In turn, 
this is the reason for which the changes in entanglement and in correlations
(when switching from positive to negative potentials) are more relevant
at low interaction energies.
A similar line of reasoning does not hold for $\left|\Psi_{5}\right\rangle $,
which is composed by the scattering states $\left|14,16\right\rangle _{s}$
and $\left|14,17\right\rangle _{s}$, both belonging to the main subband:
as we can see from Fig. \ref{fig:Delta_V_projs}, the differences
in the projections $\Delta(V)$ for $\left|\Psi_{5}\right\rangle $
are nearly constant when we increase $V$. Indeed, we do not observe
a significant change either in entanglement or in the correlation
maps at increasing $V$, as we see in Figs. \ref{fig:PsiEnta} and \ref{fig:Psi5-2-20}.
\section{Conclusions\label{sec:Conclusions}}
In this paper, we have shown that in the dynamics of two 
identical bosons evolving according to the Hubbard Hamiltonian both 
two-site correlations and entanglement of particles are differently 
affected by the sign of the on-site interaction. These
differences are more significant for low values of the interaction
$V$, i.e. of the magnitude of the hopping amplitude $J$, thus making
this regime achievable using current technologies
\cite{winkler2006repulsively,wang2010pairquench}. This behaviour
arises from the fact that the projections of some localised number states on
the eigenfunctions of the Hubbard model may change sign when we change
the sign of $V$. Therefore, the effect of $\mathrm{sgn}(V)$ may
be observed in the evolution of linear superposition of number states. 
This also explains why these features have not been pointed out and 
observed before, since most of the past literature was mainly
focused on the study of single number states.
\par
The quantitative dependence of correlations on the sign of $V$ may be 
relevant, up to a factor 2 for entanglement. This 
phenomenon also provides a further degree of freedom for manipulating 
the correlations in a quantum walk, and may be exploited to perform 
specific tasks in quantum information processing.
\par
Besides revealing novel features of the Hubbard model, and providing a 
probing technique for the sign of interaction, our results pave the way 
to further research, e.g. aimed at investigating whether this behaviour may be observed in some extensions of the Hubbard model - e.g. the Fermi-Hubbard 
model for spinless fermions
(the so-called fermion-polaron model \cite{scott1994quantum}) - or
if it is an exclusive feature of bosonic Hubbard models. Indeed, some 
recent works \cite{chattaraj2016long} have shown
a breaking of the symmetry $\pm V$ for long-range hopping in hard-core
bosons. Besides, it seems worth to explore the signatures of long-range
hopping and interactions (i.e., extended to first- and second-nearest
neighbouring sites) on the dynamics and the entanglement of both bosonic
and fermionic particles.
\acknowledgments
{This work has been supported by UniMORE through FAR2014 and by 
EU through the collaborative H2020 project QuProCS (Grant Agreement 
641277). The authors thank Ilaria Siloi for fruitful discussions. 
PB and MGAP are members of INdAM. }
\appendix
\section{Invariance under Boost and Time-reversal transformations}
In this section we want to further discuss the symmetries of the 
Hamiltonians $ { {H}}_{\pm}= { {H}}_{N}(J,\pm V)$
\cite{schneider2012fermionic,Lee2014} in order to better appreciate
the results presented in the body of the paper. In particular, we want
to emphasise that the expectation value of an operator $ {O}$ on state 
$\left|\psi(t)\right\rangle _{\pm}=\exp(-i {H}_{\pm}t)\left|\psi_{0}\right\rangle $
is independent from the sign of $V$ if both the initial state 
$\left|\psi_{0}\right\rangle $
and $ {O}$ are invariant under both \emph{boost transformation}
and \emph{time reversal transformation}. In order to better illustrate
this features we will go through the explicit proof of the 
invariance.
\par
The idempotent Hermitian \emph{boost operator}, $ {B}= {B}^{\dagger}$,
$ {B}^{2}=1$, is defined by its action in $k$ space: 
$ {B} {a}_{k}^{\dagger} {B}= {a}_{k+\pi}^{\dagger}$, whereas
in the in $x$ space we have $ {B} {c}_{j} {B}=
e^{\mathrm{-i}\pi j} {c}_{j}$. As a consequence, we may write
\begin{align}
 {B} { {H}}_{\pm} {B} = - { {H}}_{\mp}.
\end{align}
Being $B$ a unitary operator any eigenstate of $ {B}$ is actually 
invariant under the action of $ {B}$. The same discussion does not hold, 
of course, for combinations of eigenstats belonging to different eigenvalues.
The \emph{time reversal operator} is the antiunitary operator ${\Theta}$ defined
by $ {\Theta}e^{-\mathrm{i}H_{\pm}t} {\Theta}^{\dagger}=e^{+\mathrm{i}H_{\pm}t}$
and $ {\Theta}\left|\psi_{0}\right\rangle =\left|\psi_{0}^{*}\right\rangle $
and which may be written as $ {\Theta}=KU$, where $U$ is a unitary operator 
and $K$ is
the complex-conjugation operator. An operator is said to 
have a well defined symmetry under
time reversal if it is even (invariant) or odd, 
$ {\Theta} {O} {\Theta}^{\dagger}=\pm {O}$.
\par
Let us consider now the expectation value of $ {O}$ under 
$ { {H}}_{+}$, i.e.  $\left\langle O(t)\right\rangle _{+}=
\left\langle \psi_{0}\right|\exp(+i { {H}}_{+}t) {O}\exp(-i 
{ {H}}_{+}t)\left|\psi_{0}\right\rangle$.
Under the hypotesys that both $ {O}$ and $\left|\psi_{0}\right\rangle $
are invariant under $ {B}$ we have 
\begin{align}
\left\langle O(t)\right\rangle _{+} 
 & =  \left\langle \psi_{0}\right|\exp(-i { {H}}_{-}t) {O}\exp(+i { {H}}_{-}t)\left|\psi_{0}\right\rangle \notag \\
 & =  \left\langle O(-t)\right\rangle _{-}.
\end{align}
If we also consider the invariance under time reversal of both $ {O}$
and $\left|\psi_{0}\right\rangle $, we get finally
\begin{align}
\left\langle O(t)\right\rangle _{+} & =  \left\langle O(-t)\right\rangle _{-} \notag \\
 & =  \left\langle \psi_{0}\right|\exp(+i { {H}}_{-}t) {O}\exp(-i { {H}}_{-}t)\left|\psi_{0}\right\rangle \notag \\
 & =  \left\langle O(t)\right\rangle _{-},
\end{align}
so that the switching of the sign of the potential produces no effect
on the expectation value of $ {O}$. Overall, we may conclude that in 
general, if the state and the observable are invariant under boost 
transformation, when we switch the sign of the potential $V$ we actually 
reverse the direction of time: some observables, like the entanglement, 
are not affected by this operations, while others do. In particular, 
if we look at the correlation maps, changing the direction of time 
is equal to changing sign to all the velocity components of the 
wave-functions, so that left and right directions are reversed 
(i.e., the correlation maps become specular). However, this 
phenomenon is not observable if the velocity composition 
is symmetrical (i.e., it determines a symmetrical expansion on the
map), but in the opposite case it can be spotted. This is particulary
apparent for states like $\left|\psi\right\rangle =
\left|14,16\right\rangle _{s}+\left|15,17\right\rangle _{s}$,
where entanglement and correlation maps are identical (both states
have the same eigenvalue with respect to $ {B}$). On the other hand, 
if we add a complex relative phase, i.e. $\left|\psi\right\rangle =
\left|14,16\right\rangle _{s}+i\left|15,17\right\rangle _{s}$,
we break the time-reversal invariance and we make the velocity composition
asymmetrical. In this case, the entanglement does not change
its dynamics, but the correlation maps are reversed with respect 
to their anti-diagonal.
\bibliography{phbib}

\begin{thebibliography}{78}
\expandafter\ifx\csname natexlab\endcsname\relax\def\natexlab#1{#1}\fi
\expandafter\ifx\csname bibnamefont\endcsname\relax
  \def\bibnamefont#1{#1}\fi
\expandafter\ifx\csname bibfnamefont\endcsname\relax
  \def\bibfnamefont#1{#1}\fi
\expandafter\ifx\csname citenamefont\endcsname\relax
  \def\citenamefont#1{#1}\fi
\expandafter\ifx\csname url\endcsname\relax
  \def\url#1{\texttt{#1}}\fi
\expandafter\ifx\csname urlprefix\endcsname\relax\def\urlprefix{URL }\fi
\providecommand{\bibinfo}[2]{#2}
\providecommand{\eprint}[2][]{\url{#2}}

\bibitem[{\citenamefont{Essler et~al.}(2005)\citenamefont{Essler, Frahm,
  G{\"o}hmann, Kl{\"u}mper, and Korepin}}]{essler}
\bibinfo{author}{\bibfnamefont{F.~H.} \bibnamefont{Essler}},
  \bibinfo{author}{\bibfnamefont{H.}~\bibnamefont{Frahm}},
  \bibinfo{author}{\bibfnamefont{F.}~\bibnamefont{G{\"o}hmann}},
  \bibinfo{author}{\bibfnamefont{A.}~\bibnamefont{Kl{\"u}mper}},
  \bibnamefont{and} \bibinfo{author}{\bibfnamefont{V.~E.}
  \bibnamefont{Korepin}}, \emph{\bibinfo{title}{The one-dimensional Hubbard
  model}} (\bibinfo{publisher}{Cambridge University Press},
  \bibinfo{year}{2005}).

\bibitem[{\citenamefont{Montorsi}(1992)}]{montorsi1992hubbard}
\bibinfo{author}{\bibfnamefont{A.}~\bibnamefont{Montorsi}},
  \emph{\bibinfo{title}{The Hubbard Model: A Reprint Volume}}
  (\bibinfo{publisher}{World Scientific}, \bibinfo{year}{1992}).

\bibitem[{\citenamefont{Lieb and Wu}(2003)}]{lieb2003one}
\bibinfo{author}{\bibfnamefont{E.~H.} \bibnamefont{Lieb}} \bibnamefont{and}
  \bibinfo{author}{\bibfnamefont{F.}~\bibnamefont{Wu}},
  \bibinfo{journal}{Physica A} \textbf{\bibinfo{volume}{321}},
  \bibinfo{pages}{1} (\bibinfo{year}{2003}).

\bibitem[{\citenamefont{Amico et~al.}(2008)\citenamefont{Amico, Fazio,
  Osterloh, and Vedral}}]{Amico2008}
\bibinfo{author}{\bibfnamefont{L.}~\bibnamefont{Amico}},
  \bibinfo{author}{\bibfnamefont{R.}~\bibnamefont{Fazio}},
  \bibinfo{author}{\bibfnamefont{A.}~\bibnamefont{Osterloh}}, \bibnamefont{and}
  \bibinfo{author}{\bibfnamefont{V.}~\bibnamefont{Vedral}},
  \bibinfo{journal}{Rev. Mod. Phys.} \textbf{\bibinfo{volume}{80}},
  \bibinfo{pages}{517} (\bibinfo{year}{2008}).

\bibitem[{\citenamefont{Preiss et~al.}(2015)\citenamefont{Preiss, Ma, Tai,
  Lukin, Rispoli, Zupancic, Lahini, Islam, and Greiner}}]{Preiss2015}
\bibinfo{author}{\bibfnamefont{P.~M.} \bibnamefont{Preiss}},
  \bibinfo{author}{\bibfnamefont{R.}~\bibnamefont{Ma}},
  \bibinfo{author}{\bibfnamefont{M.~E.} \bibnamefont{Tai}},
  \bibinfo{author}{\bibfnamefont{A.}~\bibnamefont{Lukin}},
  \bibinfo{author}{\bibfnamefont{M.}~\bibnamefont{Rispoli}},
  \bibinfo{author}{\bibfnamefont{P.}~\bibnamefont{Zupancic}},
  \bibinfo{author}{\bibfnamefont{Y.}~\bibnamefont{Lahini}},
  \bibinfo{author}{\bibfnamefont{R.}~\bibnamefont{Islam}}, \bibnamefont{and}
  \bibinfo{author}{\bibfnamefont{M.}~\bibnamefont{Greiner}},
  \bibinfo{journal}{Science} \textbf{\bibinfo{volume}{347}},
  \bibinfo{pages}{1229} (\bibinfo{year}{2015}).

\bibitem[{\citenamefont{Duan et~al.}(2003)\citenamefont{Duan, Demler, and
  Lukin}}]{Duan2003}
\bibinfo{author}{\bibfnamefont{L.-M.} \bibnamefont{Duan}},
  \bibinfo{author}{\bibfnamefont{E.}~\bibnamefont{Demler}}, \bibnamefont{and}
  \bibinfo{author}{\bibfnamefont{M.~D.} \bibnamefont{Lukin}},
  \bibinfo{journal}{Phys. Rev. Lett.} \textbf{\bibinfo{volume}{91}},
  \bibinfo{pages}{090402} (\bibinfo{year}{2003}).

\bibitem[{\citenamefont{Fukuhara
  et~al.}(2013{\natexlab{a}})\citenamefont{Fukuhara, Schau{\ss}, Endres, Hild,
  Cheneau, Bloch, and Gross}}]{fukuhara2013}
\bibinfo{author}{\bibfnamefont{T.}~\bibnamefont{Fukuhara}},
  \bibinfo{author}{\bibfnamefont{P.}~\bibnamefont{Schau{\ss}}},
  \bibinfo{author}{\bibfnamefont{M.}~\bibnamefont{Endres}},
  \bibinfo{author}{\bibfnamefont{S.}~\bibnamefont{Hild}},
  \bibinfo{author}{\bibfnamefont{M.}~\bibnamefont{Cheneau}},
  \bibinfo{author}{\bibfnamefont{I.}~\bibnamefont{Bloch}}, \bibnamefont{and}
  \bibinfo{author}{\bibfnamefont{C.}~\bibnamefont{Gross}},
  \bibinfo{journal}{Nature} \textbf{\bibinfo{volume}{502}}, \bibinfo{pages}{76}
  (\bibinfo{year}{2013}{\natexlab{a}}).

\bibitem[{\citenamefont{Jaksch and Zoller}(2005)}]{Jaksch200552}
\bibinfo{author}{\bibfnamefont{D.}~\bibnamefont{Jaksch}} \bibnamefont{and}
  \bibinfo{author}{\bibfnamefont{P.}~\bibnamefont{Zoller}},
  \bibinfo{journal}{Ann. Phys.} \textbf{\bibinfo{volume}{315}},
  \bibinfo{pages}{52 } (\bibinfo{year}{2005}), \bibinfo{note}{special Issue}.

\bibitem[{\citenamefont{Fukuhara
  et~al.}(2013{\natexlab{b}})\citenamefont{Fukuhara, Kantian, Endres, Cheneau,
  Schau{\ss}, Hild, Bellem, Schollw{\"o}ck, Giamarchi, Gross
  et~al.}}]{fukuhara2013b}
\bibinfo{author}{\bibfnamefont{T.}~\bibnamefont{Fukuhara}},
  \bibinfo{author}{\bibfnamefont{A.}~\bibnamefont{Kantian}},
  \bibinfo{author}{\bibfnamefont{M.}~\bibnamefont{Endres}},
  \bibinfo{author}{\bibfnamefont{M.}~\bibnamefont{Cheneau}},
  \bibinfo{author}{\bibfnamefont{P.}~\bibnamefont{Schau{\ss}}},
  \bibinfo{author}{\bibfnamefont{S.}~\bibnamefont{Hild}},
  \bibinfo{author}{\bibfnamefont{D.}~\bibnamefont{Bellem}},
  \bibinfo{author}{\bibfnamefont{U.}~\bibnamefont{Schollw{\"o}ck}},
  \bibinfo{author}{\bibfnamefont{T.}~\bibnamefont{Giamarchi}},
  \bibinfo{author}{\bibfnamefont{C.}~\bibnamefont{Gross}},
  \bibnamefont{et~al.}, \bibinfo{journal}{Nature Phys.}
  \textbf{\bibinfo{volume}{9}}, \bibinfo{pages}{235}
  (\bibinfo{year}{2013}{\natexlab{b}}).

\bibitem[{\citenamefont{Schulz}(1990)}]{Schulz1990}
\bibinfo{author}{\bibfnamefont{H.~J.} \bibnamefont{Schulz}},
  \bibinfo{journal}{Phys. Rev. Lett.} \textbf{\bibinfo{volume}{64}},
  \bibinfo{pages}{2831} (\bibinfo{year}{1990}).

\bibitem[{\citenamefont{Lahini et~al.}(2012)\citenamefont{Lahini, Verbin,
  Huber, Bromberg, Pugatch, and Silberberg}}]{Lahini2012}
\bibinfo{author}{\bibfnamefont{Y.}~\bibnamefont{Lahini}},
  \bibinfo{author}{\bibfnamefont{M.}~\bibnamefont{Verbin}},
  \bibinfo{author}{\bibfnamefont{S.~D.} \bibnamefont{Huber}},
  \bibinfo{author}{\bibfnamefont{Y.}~\bibnamefont{Bromberg}},
  \bibinfo{author}{\bibfnamefont{R.}~\bibnamefont{Pugatch}}, \bibnamefont{and}
  \bibinfo{author}{\bibfnamefont{Y.}~\bibnamefont{Silberberg}},
  \bibinfo{journal}{Phys. Rev. A} \textbf{\bibinfo{volume}{86}},
  \bibinfo{pages}{011603} (\bibinfo{year}{2012}).

\bibitem[{\citenamefont{Bromberg et~al.}(2010)\citenamefont{Bromberg, Lahini,
  Small, and Silberberg}}]{Bromberg2010}
\bibinfo{author}{\bibfnamefont{Y.}~\bibnamefont{Bromberg}},
  \bibinfo{author}{\bibfnamefont{Y.}~\bibnamefont{Lahini}},
  \bibinfo{author}{\bibfnamefont{E.}~\bibnamefont{Small}}, \bibnamefont{and}
  \bibinfo{author}{\bibfnamefont{Y.}~\bibnamefont{Silberberg}},
  \bibinfo{journal}{Nature Phot.} \textbf{\bibinfo{volume}{4}},
  \bibinfo{pages}{721} (\bibinfo{year}{2010}).

\bibitem[{\citenamefont{Lederer et~al.}(2008)\citenamefont{Lederer, Stegeman,
  Christodoulides, Assanto, Segev, and Silberberg}}]{Lederer2008}
\bibinfo{author}{\bibfnamefont{F.}~\bibnamefont{Lederer}},
  \bibinfo{author}{\bibfnamefont{G.~I.} \bibnamefont{Stegeman}},
  \bibinfo{author}{\bibfnamefont{D.~N.} \bibnamefont{Christodoulides}},
  \bibinfo{author}{\bibfnamefont{G.}~\bibnamefont{Assanto}},
  \bibinfo{author}{\bibfnamefont{M.}~\bibnamefont{Segev}}, \bibnamefont{and}
  \bibinfo{author}{\bibfnamefont{Y.}~\bibnamefont{Silberberg}},
  \bibinfo{journal}{Physics Reports} \textbf{\bibinfo{volume}{463}},
  \bibinfo{pages}{1 } (\bibinfo{year}{2008}).

\bibitem[{\citenamefont{Lee et~al.}(2014)\citenamefont{Lee, Rai, Noh, and
  Angelakis}}]{Lee2014}
\bibinfo{author}{\bibfnamefont{C.}~\bibnamefont{Lee}},
  \bibinfo{author}{\bibfnamefont{A.}~\bibnamefont{Rai}},
  \bibinfo{author}{\bibfnamefont{C.}~\bibnamefont{Noh}}, \bibnamefont{and}
  \bibinfo{author}{\bibfnamefont{D.~G.} \bibnamefont{Angelakis}},
  \bibinfo{journal}{Phys. Rev. A} \textbf{\bibinfo{volume}{89}},
  \bibinfo{pages}{023823} (\bibinfo{year}{2014}).

\bibitem[{\citenamefont{Longhi}(2011)}]{longhi2011optical}
\bibinfo{author}{\bibfnamefont{S.}~\bibnamefont{Longhi}}, \bibinfo{journal}{J.
  Phys. B} \textbf{\bibinfo{volume}{44}}, \bibinfo{pages}{051001}
  (\bibinfo{year}{2011}).

\bibitem[{\citenamefont{Matsubara and Matsuda}(1956)}]{matsubara1956lattice}
\bibinfo{author}{\bibfnamefont{T.}~\bibnamefont{Matsubara}} \bibnamefont{and}
  \bibinfo{author}{\bibfnamefont{H.}~\bibnamefont{Matsuda}},
  \bibinfo{journal}{Progr. Theor. Phys.} \textbf{\bibinfo{volume}{16}},
  \bibinfo{pages}{569} (\bibinfo{year}{1956}).

\bibitem[{\citenamefont{Qin et~al.}(2014)\citenamefont{Qin, Ke, Guan, Li,
  Andrei, and Lee}}]{Zong-Qin2014}
\bibinfo{author}{\bibfnamefont{X.}~\bibnamefont{Qin}},
  \bibinfo{author}{\bibfnamefont{Y.}~\bibnamefont{Ke}},
  \bibinfo{author}{\bibfnamefont{X.}~\bibnamefont{Guan}},
  \bibinfo{author}{\bibfnamefont{Z.}~\bibnamefont{Li}},
  \bibinfo{author}{\bibfnamefont{N.}~\bibnamefont{Andrei}}, \bibnamefont{and}
  \bibinfo{author}{\bibfnamefont{C.}~\bibnamefont{Lee}},
  \bibinfo{journal}{Phys. Rev. A} \textbf{\bibinfo{volume}{90}},
  \bibinfo{pages}{062301} (\bibinfo{year}{2014}).

\bibitem[{\citenamefont{Wang et~al.}(2014)\citenamefont{Wang, Wang, and
  Zhang}}]{wang2014anyons}
\bibinfo{author}{\bibfnamefont{L.}~\bibnamefont{Wang}},
  \bibinfo{author}{\bibfnamefont{L.}~\bibnamefont{Wang}}, \bibnamefont{and}
  \bibinfo{author}{\bibfnamefont{Y.}~\bibnamefont{Zhang}},
  \bibinfo{journal}{Phys. Rev. A} \textbf{\bibinfo{volume}{90}},
  \bibinfo{pages}{063618} (\bibinfo{year}{2014}).

\bibitem[{\citenamefont{Wang et~al.}(2015)\citenamefont{Wang, Liu, Chen, and
  Zhang}}]{Wang2015spinflipQW}
\bibinfo{author}{\bibfnamefont{L.}~\bibnamefont{Wang}},
  \bibinfo{author}{\bibfnamefont{N.}~\bibnamefont{Liu}},
  \bibinfo{author}{\bibfnamefont{S.}~\bibnamefont{Chen}}, \bibnamefont{and}
  \bibinfo{author}{\bibfnamefont{Y.}~\bibnamefont{Zhang}},
  \bibinfo{journal}{Phys. Rev. A} \textbf{\bibinfo{volume}{92}},
  \bibinfo{pages}{053606} (\bibinfo{year}{2015}).

\bibitem[{\citenamefont{Hecker~Denschlag and Daley}(2007)}]{Denschlag2007}
\bibinfo{author}{\bibfnamefont{J.}~\bibnamefont{Hecker~Denschlag}}
  \bibnamefont{and} \bibinfo{author}{\bibfnamefont{A.~J.} \bibnamefont{Daley}},
  in \emph{\bibinfo{booktitle}{Proceedings of the International School of
  Physics "Enrico Fermi"}} (\bibinfo{publisher}{IOS Press Ebooks},
  \bibinfo{year}{2007}), vol. \bibinfo{volume}{164}.

\bibitem[{\citenamefont{Winkler et~al.}(2006)\citenamefont{Winkler, Thalhammer,
  Lang, Grimm, Denschlag, Daley, Kantian, B{\"u}chler, and
  Zoller}}]{winkler2006repulsively}
\bibinfo{author}{\bibfnamefont{K.}~\bibnamefont{Winkler}},
  \bibinfo{author}{\bibfnamefont{G.}~\bibnamefont{Thalhammer}},
  \bibinfo{author}{\bibfnamefont{F.}~\bibnamefont{Lang}},
  \bibinfo{author}{\bibfnamefont{R.}~\bibnamefont{Grimm}},
  \bibinfo{author}{\bibfnamefont{J.~H.} \bibnamefont{Denschlag}},
  \bibinfo{author}{\bibfnamefont{A.}~\bibnamefont{Daley}},
  \bibinfo{author}{\bibfnamefont{A.}~\bibnamefont{Kantian}},
  \bibinfo{author}{\bibfnamefont{H.}~\bibnamefont{B{\"u}chler}},
  \bibnamefont{and} \bibinfo{author}{\bibfnamefont{P.}~\bibnamefont{Zoller}},
  \bibinfo{journal}{Nature} \textbf{\bibinfo{volume}{441}},
  \bibinfo{pages}{853} (\bibinfo{year}{2006}).

\bibitem[{\citenamefont{Wang et~al.}(2010)\citenamefont{Wang, Hao, and
  Chen}}]{wang2010pairquench}
\bibinfo{author}{\bibfnamefont{L.}~\bibnamefont{Wang}},
  \bibinfo{author}{\bibfnamefont{Y.}~\bibnamefont{Hao}}, \bibnamefont{and}
  \bibinfo{author}{\bibfnamefont{S.}~\bibnamefont{Chen}},
  \bibinfo{journal}{Phys. Rev. A} \textbf{\bibinfo{volume}{81}},
  \bibinfo{pages}{063637} (\bibinfo{year}{2010}).

\bibitem[{\citenamefont{F{\"o}lling et~al.}(2007)\citenamefont{F{\"o}lling,
  Trotzky, Cheinet, Feld, Saers, Widera, M{\"u}ller, and
  Bloch}}]{folling2007direct}
\bibinfo{author}{\bibfnamefont{S.}~\bibnamefont{F{\"o}lling}},
  \bibinfo{author}{\bibfnamefont{S.}~\bibnamefont{Trotzky}},
  \bibinfo{author}{\bibfnamefont{P.}~\bibnamefont{Cheinet}},
  \bibinfo{author}{\bibfnamefont{M.}~\bibnamefont{Feld}},
  \bibinfo{author}{\bibfnamefont{R.}~\bibnamefont{Saers}},
  \bibinfo{author}{\bibfnamefont{A.}~\bibnamefont{Widera}},
  \bibinfo{author}{\bibfnamefont{T.}~\bibnamefont{M{\"u}ller}},
  \bibnamefont{and} \bibinfo{author}{\bibfnamefont{I.}~\bibnamefont{Bloch}},
  \bibinfo{journal}{Nature} \textbf{\bibinfo{volume}{448}},
  \bibinfo{pages}{1029} (\bibinfo{year}{2007}).

\bibitem[{\citenamefont{Piil and M\o{}lmer}(2007)}]{Piil2007}
\bibinfo{author}{\bibfnamefont{R.}~\bibnamefont{Piil}} \bibnamefont{and}
  \bibinfo{author}{\bibfnamefont{K.}~\bibnamefont{M\o{}lmer}},
  \bibinfo{journal}{Phys. Rev. A} \textbf{\bibinfo{volume}{76}},
  \bibinfo{pages}{023607} (\bibinfo{year}{2007}).

\bibitem[{\citenamefont{Wang et~al.}(2008)\citenamefont{Wang, Hao, and
  Chen}}]{wang2008quantum}
\bibinfo{author}{\bibfnamefont{L.}~\bibnamefont{Wang}},
  \bibinfo{author}{\bibfnamefont{Y.}~\bibnamefont{Hao}}, \bibnamefont{and}
  \bibinfo{author}{\bibfnamefont{S.}~\bibnamefont{Chen}},
  \bibinfo{journal}{Eur. Phys. J. D} \textbf{\bibinfo{volume}{48}},
  \bibinfo{pages}{229} (\bibinfo{year}{2008}).

\bibitem[{\citenamefont{Valiente and Petrosyan}(2008)}]{valiente2008two}
\bibinfo{author}{\bibfnamefont{M.}~\bibnamefont{Valiente}} \bibnamefont{and}
  \bibinfo{author}{\bibfnamefont{D.}~\bibnamefont{Petrosyan}},
  \bibinfo{journal}{J. Phys. B} \textbf{\bibinfo{volume}{41}},
  \bibinfo{pages}{161002} (\bibinfo{year}{2008}).

\bibitem[{\citenamefont{Schneider et~al.}(2012)\citenamefont{Schneider,
  Hackerm{\"u}ller, Ronzheimer, Will, Braun, Best, Bloch, Demler, Mandt, Rasch
  et~al.}}]{schneider2012fermionic}
\bibinfo{author}{\bibfnamefont{U.}~\bibnamefont{Schneider}},
  \bibinfo{author}{\bibfnamefont{L.}~\bibnamefont{Hackerm{\"u}ller}},
  \bibinfo{author}{\bibfnamefont{J.~P.} \bibnamefont{Ronzheimer}},
  \bibinfo{author}{\bibfnamefont{S.}~\bibnamefont{Will}},
  \bibinfo{author}{\bibfnamefont{S.}~\bibnamefont{Braun}},
  \bibinfo{author}{\bibfnamefont{T.}~\bibnamefont{Best}},
  \bibinfo{author}{\bibfnamefont{I.}~\bibnamefont{Bloch}},
  \bibinfo{author}{\bibfnamefont{E.}~\bibnamefont{Demler}},
  \bibinfo{author}{\bibfnamefont{S.}~\bibnamefont{Mandt}},
  \bibinfo{author}{\bibfnamefont{D.}~\bibnamefont{Rasch}},
  \bibnamefont{et~al.}, \bibinfo{journal}{Nature Phys.}
  \textbf{\bibinfo{volume}{8}}, \bibinfo{pages}{213} (\bibinfo{year}{2012}).

\bibitem[{\citenamefont{Beggi et~al.}(2016)\citenamefont{Beggi, Buscemi, and
  Bordone}}]{Beggi2016colonoise}
\bibinfo{author}{\bibfnamefont{A.}~\bibnamefont{Beggi}},
  \bibinfo{author}{\bibfnamefont{F.}~\bibnamefont{Buscemi}}, \bibnamefont{and}
  \bibinfo{author}{\bibfnamefont{P.}~\bibnamefont{Bordone}},
  \bibinfo{journal}{Quantum Inf. Proc.} \textbf{\bibinfo{volume}{15}},
  \bibinfo{pages}{3711} (\bibinfo{year}{2016}).

\bibitem[{\citenamefont{Benedetti et~al.}(2014)\citenamefont{Benedetti,
  Buscemi, Bordone, and Paris}}]{qpsp14}
\bibinfo{author}{\bibfnamefont{C.}~\bibnamefont{Benedetti}},
  \bibinfo{author}{\bibfnamefont{F.}~\bibnamefont{Buscemi}},
  \bibinfo{author}{\bibfnamefont{P.}~\bibnamefont{Bordone}}, \bibnamefont{and}
  \bibinfo{author}{\bibfnamefont{M.~G.~A.} \bibnamefont{Paris}},
  \bibinfo{journal}{Phys. Rev. A} \textbf{\bibinfo{volume}{89}},
  \bibinfo{pages}{032114} (\bibinfo{year}{2014}).

\bibitem[{\citenamefont{Paris}(2014)}]{frac14}
\bibinfo{author}{\bibfnamefont{M.~G.~A.} \bibnamefont{Paris}},
  \bibinfo{journal}{Physica A} \textbf{\bibinfo{volume}{413}},
  \bibinfo{pages}{256 } (\bibinfo{year}{2014}).

\bibitem[{\citenamefont{Tamascelli
  et~al.}(2016{\natexlab{a}})\citenamefont{Tamascelli, Benedetti, Olivares, and
  Paris}}]{fprobes}
\bibinfo{author}{\bibfnamefont{D.}~\bibnamefont{Tamascelli}},
  \bibinfo{author}{\bibfnamefont{C.}~\bibnamefont{Benedetti}},
  \bibinfo{author}{\bibfnamefont{S.}~\bibnamefont{Olivares}}, \bibnamefont{and}
  \bibinfo{author}{\bibfnamefont{M.~G.~A.} \bibnamefont{Paris}},
  \bibinfo{journal}{Phys. Rev. A} \textbf{\bibinfo{volume}{94}},
  \bibinfo{pages}{042129} (\bibinfo{year}{2016}{\natexlab{a}}).

\bibitem[{\citenamefont{Siloi et~al.}(2017)\citenamefont{Siloi, Benedetti,
  Piccinini, Piilo, Maniscalco, Paris, and Bordone}}]{siloi17}
\bibinfo{author}{\bibfnamefont{I.}~\bibnamefont{Siloi}},
  \bibinfo{author}{\bibfnamefont{C.}~\bibnamefont{Benedetti}},
  \bibinfo{author}{\bibfnamefont{E.}~\bibnamefont{Piccinini}},
  \bibinfo{author}{\bibfnamefont{J.}~\bibnamefont{Piilo}},
  \bibinfo{author}{\bibfnamefont{S.}~\bibnamefont{Maniscalco}},
  \bibinfo{author}{\bibfnamefont{M.~G.~A.} \bibnamefont{Paris}},
  \bibnamefont{and} \bibinfo{author}{\bibfnamefont{P.}~\bibnamefont{Bordone}},
  \bibinfo{journal}{Phys. Rev. A} \textbf{\bibinfo{volume}{95}},
  \bibinfo{pages}{022106} (\bibinfo{year}{2017}).

\bibitem[{\citenamefont{Kempe}(2003)}]{Kempe2003_QW}
\bibinfo{author}{\bibfnamefont{J.}~\bibnamefont{Kempe}},
  \bibinfo{journal}{Contemp. Phys.} \textbf{\bibinfo{volume}{44}},
  \bibinfo{pages}{307} (\bibinfo{year}{2003}).

\bibitem[{\citenamefont{Venegas-Andraca}(2012)}]{Venegas2012_QW}
\bibinfo{author}{\bibfnamefont{S.}~\bibnamefont{Venegas-Andraca}},
  \bibinfo{journal}{Quantum Inf. Proc.} \textbf{\bibinfo{volume}{11}},
  \bibinfo{pages}{1015} (\bibinfo{year}{2012}).

\bibitem[{\citenamefont{Ambainis}(2003)}]{Ambainis2003}
\bibinfo{author}{\bibfnamefont{A.}~\bibnamefont{Ambainis}},
  \bibinfo{journal}{Int. J. Quantum Inf.} \textbf{\bibinfo{volume}{01}},
  \bibinfo{pages}{507} (\bibinfo{year}{2003}).

\bibitem[{\citenamefont{Portugal}(2013)}]{portugal2013quantum}
\bibinfo{author}{\bibfnamefont{R.}~\bibnamefont{Portugal}},
  \emph{\bibinfo{title}{Quantum walks and search algorithms}}
  (\bibinfo{publisher}{Springer Science \& Business Media},
  \bibinfo{year}{2013}).

\bibitem[{\citenamefont{Venegas-Andraca}(2008)}]{venegas2008quantum}
\bibinfo{author}{\bibfnamefont{S.~E.} \bibnamefont{Venegas-Andraca}},
  \bibinfo{journal}{Synthesis Lectures on Quantum Computing}
  \textbf{\bibinfo{volume}{1}}, \bibinfo{pages}{1} (\bibinfo{year}{2008}).

\bibitem[{\citenamefont{Tamascelli and Zanetti}(2014)}]{dt14}
\bibinfo{author}{\bibfnamefont{D.}~\bibnamefont{Tamascelli}} \bibnamefont{and}
  \bibinfo{author}{\bibfnamefont{L.}~\bibnamefont{Zanetti}},
  \bibinfo{journal}{J.Phys. A} \textbf{\bibinfo{volume}{47}},
  \bibinfo{pages}{325302} (\bibinfo{year}{2014}).

\bibitem[{\citenamefont{Bromberg et~al.}(2009)\citenamefont{Bromberg, Lahini,
  Morandotti, and Silberberg}}]{Bromberg2009}
\bibinfo{author}{\bibfnamefont{Y.}~\bibnamefont{Bromberg}},
  \bibinfo{author}{\bibfnamefont{Y.}~\bibnamefont{Lahini}},
  \bibinfo{author}{\bibfnamefont{R.}~\bibnamefont{Morandotti}},
  \bibnamefont{and}
  \bibinfo{author}{\bibfnamefont{Y.}~\bibnamefont{Silberberg}},
  \bibinfo{journal}{Phys. Rev. Lett.} \textbf{\bibinfo{volume}{102}},
  \bibinfo{pages}{253904} (\bibinfo{year}{2009}).

\bibitem[{\citenamefont{Peruzzo et~al.}(2010)\citenamefont{Peruzzo, Lobino,
  Matthews, Matsuda, Politi, Poulios, Zhou, Lahini, Ismail, Wörhoff
  et~al.}}]{Peruzzo2010}
\bibinfo{author}{\bibfnamefont{A.}~\bibnamefont{Peruzzo}},
  \bibinfo{author}{\bibfnamefont{M.}~\bibnamefont{Lobino}},
  \bibinfo{author}{\bibfnamefont{J.~C.~F.} \bibnamefont{Matthews}},
  \bibinfo{author}{\bibfnamefont{N.}~\bibnamefont{Matsuda}},
  \bibinfo{author}{\bibfnamefont{A.}~\bibnamefont{Politi}},
  \bibinfo{author}{\bibfnamefont{K.}~\bibnamefont{Poulios}},
  \bibinfo{author}{\bibfnamefont{X.-Q.} \bibnamefont{Zhou}},
  \bibinfo{author}{\bibfnamefont{Y.}~\bibnamefont{Lahini}},
  \bibinfo{author}{\bibfnamefont{N.}~\bibnamefont{Ismail}},
  \bibinfo{author}{\bibfnamefont{K.}~\bibnamefont{Wörhoff}},
  \bibnamefont{et~al.}, \bibinfo{journal}{Science}
  \textbf{\bibinfo{volume}{329}}, \bibinfo{pages}{1500} (\bibinfo{year}{2010}).

\bibitem[{\citenamefont{Rai et~al.}(2008)\citenamefont{Rai, Agarwal, and
  Perk}}]{Rai2008}
\bibinfo{author}{\bibfnamefont{A.}~\bibnamefont{Rai}},
  \bibinfo{author}{\bibfnamefont{G.~S.} \bibnamefont{Agarwal}},
  \bibnamefont{and} \bibinfo{author}{\bibfnamefont{J.~H.~H.}
  \bibnamefont{Perk}}, \bibinfo{journal}{Phys. Rev. A}
  \textbf{\bibinfo{volume}{78}}, \bibinfo{pages}{042304}
  (\bibinfo{year}{2008}).

\bibitem[{\citenamefont{Tamascelli
  et~al.}(2016{\natexlab{b}})\citenamefont{Tamascelli, Olivares, Rossotti,
  Osellame, and Paris}}]{Tamascelli2016}
\bibinfo{author}{\bibfnamefont{D.}~\bibnamefont{Tamascelli}},
  \bibinfo{author}{\bibfnamefont{S.}~\bibnamefont{Olivares}},
  \bibinfo{author}{\bibfnamefont{S.}~\bibnamefont{Rossotti}},
  \bibinfo{author}{\bibfnamefont{R.}~\bibnamefont{Osellame}}, \bibnamefont{and}
  \bibinfo{author}{\bibfnamefont{M.~G.~A.} \bibnamefont{Paris}},
  \textbf{\bibinfo{volume}{6}}, \bibinfo{pages}{26054}
  (\bibinfo{year}{2016}{\natexlab{b}}).

\bibitem[{\citenamefont{Wang and Manouchehri}(2013)}]{wang2013physical}
\bibinfo{author}{\bibfnamefont{J.}~\bibnamefont{Wang}} \bibnamefont{and}
  \bibinfo{author}{\bibfnamefont{K.}~\bibnamefont{Manouchehri}},
  \emph{\bibinfo{title}{Physical implementation of quantum walks}}
  (\bibinfo{publisher}{Springer}, \bibinfo{year}{2013}).

\bibitem[{\citenamefont{Broome et~al.}(2013)\citenamefont{Broome, Fedrizzi,
  Rahimi-Keshari, Dove, Aaronson, Ralph, and White}}]{Broome2013}
\bibinfo{author}{\bibfnamefont{M.~A.} \bibnamefont{Broome}},
  \bibinfo{author}{\bibfnamefont{A.}~\bibnamefont{Fedrizzi}},
  \bibinfo{author}{\bibfnamefont{S.}~\bibnamefont{Rahimi-Keshari}},
  \bibinfo{author}{\bibfnamefont{J.}~\bibnamefont{Dove}},
  \bibinfo{author}{\bibfnamefont{S.}~\bibnamefont{Aaronson}},
  \bibinfo{author}{\bibfnamefont{T.~C.} \bibnamefont{Ralph}}, \bibnamefont{and}
  \bibinfo{author}{\bibfnamefont{A.~G.} \bibnamefont{White}},
  \bibinfo{journal}{Science} \textbf{\bibinfo{volume}{339}},
  \bibinfo{pages}{794} (\bibinfo{year}{2013}).

\bibitem[{\citenamefont{Spring et~al.}(2013)\citenamefont{Spring, Metcalf,
  Humphreys, Kolthammer, Jin, Barbieri, Datta, Thomas-Peter, Langford, Kundys
  et~al.}}]{Spring2013}
\bibinfo{author}{\bibfnamefont{J.~B.} \bibnamefont{Spring}},
  \bibinfo{author}{\bibfnamefont{B.~J.} \bibnamefont{Metcalf}},
  \bibinfo{author}{\bibfnamefont{P.~C.} \bibnamefont{Humphreys}},
  \bibinfo{author}{\bibfnamefont{W.~S.} \bibnamefont{Kolthammer}},
  \bibinfo{author}{\bibfnamefont{X.-M.} \bibnamefont{Jin}},
  \bibinfo{author}{\bibfnamefont{M.}~\bibnamefont{Barbieri}},
  \bibinfo{author}{\bibfnamefont{A.}~\bibnamefont{Datta}},
  \bibinfo{author}{\bibfnamefont{N.}~\bibnamefont{Thomas-Peter}},
  \bibinfo{author}{\bibfnamefont{N.~K.} \bibnamefont{Langford}},
  \bibinfo{author}{\bibfnamefont{D.}~\bibnamefont{Kundys}},
  \bibnamefont{et~al.}, \bibinfo{journal}{Science}
  \textbf{\bibinfo{volume}{339}}, \bibinfo{pages}{798} (\bibinfo{year}{2013}).

\bibitem[{\citenamefont{Tillmann et~al.}(2013)\citenamefont{Tillmann,
  Daki{\'c}, Heilmann, Nolte, Szameit, and Walther}}]{Tillmann2013}
\bibinfo{author}{\bibfnamefont{M.}~\bibnamefont{Tillmann}},
  \bibinfo{author}{\bibfnamefont{B.}~\bibnamefont{Daki{\'c}}},
  \bibinfo{author}{\bibfnamefont{R.}~\bibnamefont{Heilmann}},
  \bibinfo{author}{\bibfnamefont{S.}~\bibnamefont{Nolte}},
  \bibinfo{author}{\bibfnamefont{A.}~\bibnamefont{Szameit}}, \bibnamefont{and}
  \bibinfo{author}{\bibfnamefont{P.}~\bibnamefont{Walther}},
  \bibinfo{journal}{Nature Phot.} \textbf{\bibinfo{volume}{7}},
  \bibinfo{pages}{540} (\bibinfo{year}{2013}).

\bibitem[{\citenamefont{Franson}(2013)}]{Franson2013}
\bibinfo{author}{\bibfnamefont{J.~D.} \bibnamefont{Franson}},
  \bibinfo{journal}{Science} \textbf{\bibinfo{volume}{339}},
  \bibinfo{pages}{767} (\bibinfo{year}{2013}).

\bibitem[{\citenamefont{Benedetti et~al.}(2012)\citenamefont{Benedetti,
  Buscemi, and Bordone}}]{Benedetti2012_QWferbos}
\bibinfo{author}{\bibfnamefont{C.}~\bibnamefont{Benedetti}},
  \bibinfo{author}{\bibfnamefont{F.}~\bibnamefont{Buscemi}}, \bibnamefont{and}
  \bibinfo{author}{\bibfnamefont{P.}~\bibnamefont{Bordone}},
  \bibinfo{journal}{Phys. Rev. A} \textbf{\bibinfo{volume}{85}},
  \bibinfo{pages}{042314} (\bibinfo{year}{2012}).

\bibitem[{\citenamefont{Mayer et~al.}(2011)\citenamefont{Mayer, Tichy, Mintert,
  Konrad, and Buchleitner}}]{Mayer2011}
\bibinfo{author}{\bibfnamefont{K.}~\bibnamefont{Mayer}},
  \bibinfo{author}{\bibfnamefont{M.~C.} \bibnamefont{Tichy}},
  \bibinfo{author}{\bibfnamefont{F.}~\bibnamefont{Mintert}},
  \bibinfo{author}{\bibfnamefont{T.}~\bibnamefont{Konrad}}, \bibnamefont{and}
  \bibinfo{author}{\bibfnamefont{A.}~\bibnamefont{Buchleitner}},
  \bibinfo{journal}{Phys. Rev. A} \textbf{\bibinfo{volume}{83}},
  \bibinfo{pages}{062307} (\bibinfo{year}{2011}).

\bibitem[{\citenamefont{Schliemann et~al.}(2001)\citenamefont{Schliemann,
  Cirac, Ku\ifmmode~\acute{s}\else \'{s}\fi{}, Lewenstein, and
  Loss}}]{Schliemann2001}
\bibinfo{author}{\bibfnamefont{J.}~\bibnamefont{Schliemann}},
  \bibinfo{author}{\bibfnamefont{J.~I.} \bibnamefont{Cirac}},
  \bibinfo{author}{\bibfnamefont{M.}~\bibnamefont{Ku\ifmmode~\acute{s}\else
  \'{s}\fi{}}}, \bibinfo{author}{\bibfnamefont{M.}~\bibnamefont{Lewenstein}},
  \bibnamefont{and} \bibinfo{author}{\bibfnamefont{D.}~\bibnamefont{Loss}},
  \bibinfo{journal}{Phys. Rev. A} \textbf{\bibinfo{volume}{64}},
  \bibinfo{pages}{022303} (\bibinfo{year}{2001}).

\bibitem[{\citenamefont{Eckert et~al.}(2002)\citenamefont{Eckert, Schliemann,
  Bruß, and Lewenstein}}]{Eckert2002}
\bibinfo{author}{\bibfnamefont{K.}~\bibnamefont{Eckert}},
  \bibinfo{author}{\bibfnamefont{J.}~\bibnamefont{Schliemann}},
  \bibinfo{author}{\bibfnamefont{D.}~\bibnamefont{Bruß}}, \bibnamefont{and}
  \bibinfo{author}{\bibfnamefont{M.}~\bibnamefont{Lewenstein}},
  \bibinfo{journal}{Ann. Phys.} \textbf{\bibinfo{volume}{299}},
  \bibinfo{pages}{88 } (\bibinfo{year}{2002}).

\bibitem[{\citenamefont{Buscemi et~al.}(2006)\citenamefont{Buscemi, Bordone,
  and Bertoni}}]{Buscemi2006}
\bibinfo{author}{\bibfnamefont{F.}~\bibnamefont{Buscemi}},
  \bibinfo{author}{\bibfnamefont{P.}~\bibnamefont{Bordone}}, \bibnamefont{and}
  \bibinfo{author}{\bibfnamefont{A.}~\bibnamefont{Bertoni}},
  \bibinfo{journal}{Phys. Rev. A} \textbf{\bibinfo{volume}{73}},
  \bibinfo{pages}{052312} (\bibinfo{year}{2006}).

\bibitem[{\citenamefont{Buscemi et~al.}(2007)\citenamefont{Buscemi, Bordone,
  and Bertoni}}]{Buscemi2007}
\bibinfo{author}{\bibfnamefont{F.}~\bibnamefont{Buscemi}},
  \bibinfo{author}{\bibfnamefont{P.}~\bibnamefont{Bordone}}, \bibnamefont{and}
  \bibinfo{author}{\bibfnamefont{A.}~\bibnamefont{Bertoni}},
  \bibinfo{journal}{Phys. Rev. A} \textbf{\bibinfo{volume}{75}},
  \bibinfo{pages}{032301} (\bibinfo{year}{2007}).

\bibitem[{\citenamefont{Zanardi}(2002)}]{Zanardi2002}
\bibinfo{author}{\bibfnamefont{P.}~\bibnamefont{Zanardi}},
  \bibinfo{journal}{Phys. Rev. A} \textbf{\bibinfo{volume}{65}},
  \bibinfo{pages}{042101} (\bibinfo{year}{2002}).

\bibitem[{\citenamefont{Ghirardi et~al.}(2002)\citenamefont{Ghirardi,
  Marinatto, and Weber}}]{Ghirardi2002}
\bibinfo{author}{\bibfnamefont{G.~C.} \bibnamefont{Ghirardi}},
  \bibinfo{author}{\bibfnamefont{L.}~\bibnamefont{Marinatto}},
  \bibnamefont{and} \bibinfo{author}{\bibfnamefont{T.}~\bibnamefont{Weber}},
  \bibinfo{journal}{J. Stat. Phys.} \textbf{\bibinfo{volume}{108}},
  \bibinfo{pages}{49} (\bibinfo{year}{2002}).

\bibitem[{\citenamefont{Ghirardi and Marinatto}(2004)}]{Ghirardi2004}
\bibinfo{author}{\bibfnamefont{G.~C.} \bibnamefont{Ghirardi}} \bibnamefont{and}
  \bibinfo{author}{\bibfnamefont{L.}~\bibnamefont{Marinatto}},
  \bibinfo{journal}{Phys. Rev. A} \textbf{\bibinfo{volume}{70}},
  \bibinfo{pages}{012109} (\bibinfo{year}{2004}).

\bibitem[{\citenamefont{Benatti
  et~al.}(2014{\natexlab{a}})\citenamefont{Benatti, Floreanini, and
  Titimbo}}]{Benatti2014}
\bibinfo{author}{\bibfnamefont{F.}~\bibnamefont{Benatti}},
  \bibinfo{author}{\bibfnamefont{R.}~\bibnamefont{Floreanini}},
  \bibnamefont{and} \bibinfo{author}{\bibfnamefont{K.}~\bibnamefont{Titimbo}},
  \bibinfo{journal}{Open Sys. Inf. Dyn.} \textbf{\bibinfo{volume}{21}},
  \bibinfo{pages}{1440003} (\bibinfo{year}{2014}{\natexlab{a}}).

\bibitem[{\citenamefont{Wiseman and Vaccaro}(2003)}]{Wiseman2003}
\bibinfo{author}{\bibfnamefont{H.~M.} \bibnamefont{Wiseman}} \bibnamefont{and}
  \bibinfo{author}{\bibfnamefont{J.~A.} \bibnamefont{Vaccaro}},
  \bibinfo{journal}{Phys. Rev. Lett.} \textbf{\bibinfo{volume}{91}},
  \bibinfo{pages}{097902} (\bibinfo{year}{2003}).

\bibitem[{\citenamefont{Dowling et~al.}(2006)\citenamefont{Dowling, Doherty,
  and Wiseman}}]{Dowling2006}
\bibinfo{author}{\bibfnamefont{M.~R.} \bibnamefont{Dowling}},
  \bibinfo{author}{\bibfnamefont{A.~C.} \bibnamefont{Doherty}},
  \bibnamefont{and} \bibinfo{author}{\bibfnamefont{H.~M.}
  \bibnamefont{Wiseman}}, \bibinfo{journal}{Phys. Rev. A}
  \textbf{\bibinfo{volume}{73}}, \bibinfo{pages}{052323}
  (\bibinfo{year}{2006}).

\bibitem[{\citenamefont{Sasaki et~al.}(2011)\citenamefont{Sasaki, Ichikawa, and
  Tsutsui}}]{Sasaki2011}
\bibinfo{author}{\bibfnamefont{T.}~\bibnamefont{Sasaki}},
  \bibinfo{author}{\bibfnamefont{T.}~\bibnamefont{Ichikawa}}, \bibnamefont{and}
  \bibinfo{author}{\bibfnamefont{I.}~\bibnamefont{Tsutsui}},
  \bibinfo{journal}{Phys. Rev. A} \textbf{\bibinfo{volume}{83}},
  \bibinfo{pages}{012113} (\bibinfo{year}{2011}).

\bibitem[{\citenamefont{Iemini et~al.}(2013)\citenamefont{Iemini, Maciel,
  Debarba, and Vianna}}]{Iemini2013}
\bibinfo{author}{\bibfnamefont{F.}~\bibnamefont{Iemini}},
  \bibinfo{author}{\bibfnamefont{T.}~\bibnamefont{Maciel}},
  \bibinfo{author}{\bibfnamefont{T.}~\bibnamefont{Debarba}}, \bibnamefont{and}
  \bibinfo{author}{\bibfnamefont{R.}~\bibnamefont{Vianna}},
  \bibinfo{journal}{Quantum Inf. Proc.} \textbf{\bibinfo{volume}{12}},
  \bibinfo{pages}{733} (\bibinfo{year}{2013}).

\bibitem[{\citenamefont{Iemini and Vianna}(2013)}]{Iemini_PhysRevA.87.022327}
\bibinfo{author}{\bibfnamefont{F.}~\bibnamefont{Iemini}} \bibnamefont{and}
  \bibinfo{author}{\bibfnamefont{R.~O.} \bibnamefont{Vianna}},
  \bibinfo{journal}{Phys. Rev. A} \textbf{\bibinfo{volume}{87}},
  \bibinfo{pages}{022327} (\bibinfo{year}{2013}).

\bibitem[{\citenamefont{Iemini et~al.}(2014)\citenamefont{Iemini, Debarba, and
  Vianna}}]{Iemini_PhysRevA.89.032324}
\bibinfo{author}{\bibfnamefont{F.}~\bibnamefont{Iemini}},
  \bibinfo{author}{\bibfnamefont{T.}~\bibnamefont{Debarba}}, \bibnamefont{and}
  \bibinfo{author}{\bibfnamefont{R.~O.} \bibnamefont{Vianna}},
  \bibinfo{journal}{Phys. Rev. A} \textbf{\bibinfo{volume}{89}},
  \bibinfo{pages}{032324} (\bibinfo{year}{2014}).

\bibitem[{\citenamefont{Reusch et~al.}(2015)\citenamefont{Reusch, Sperling, and
  Vogel}}]{Reusch_PhysRevA.91.042324}
\bibinfo{author}{\bibfnamefont{A.}~\bibnamefont{Reusch}},
  \bibinfo{author}{\bibfnamefont{J.}~\bibnamefont{Sperling}}, \bibnamefont{and}
  \bibinfo{author}{\bibfnamefont{W.}~\bibnamefont{Vogel}},
  \bibinfo{journal}{Phys. Rev. A} \textbf{\bibinfo{volume}{91}},
  \bibinfo{pages}{042324} (\bibinfo{year}{2015}).

\bibitem[{\citenamefont{Zanardi et~al.}(2004)\citenamefont{Zanardi, Lidar, and
  Lloyd}}]{Zanardi2004}
\bibinfo{author}{\bibfnamefont{P.}~\bibnamefont{Zanardi}},
  \bibinfo{author}{\bibfnamefont{D.~A.} \bibnamefont{Lidar}}, \bibnamefont{and}
  \bibinfo{author}{\bibfnamefont{S.}~\bibnamefont{Lloyd}},
  \bibinfo{journal}{Phys. Rev. Lett.} \textbf{\bibinfo{volume}{92}},
  \bibinfo{pages}{060402} (\bibinfo{year}{2004}).

\bibitem[{\citenamefont{Barnum et~al.}(2004)\citenamefont{Barnum, Knill, Ortiz,
  Somma, and Viola}}]{Barnum2004}
\bibinfo{author}{\bibfnamefont{H.}~\bibnamefont{Barnum}},
  \bibinfo{author}{\bibfnamefont{E.}~\bibnamefont{Knill}},
  \bibinfo{author}{\bibfnamefont{G.}~\bibnamefont{Ortiz}},
  \bibinfo{author}{\bibfnamefont{R.}~\bibnamefont{Somma}}, \bibnamefont{and}
  \bibinfo{author}{\bibfnamefont{L.}~\bibnamefont{Viola}},
  \bibinfo{journal}{Phys. Rev. Lett.} \textbf{\bibinfo{volume}{92}},
  \bibinfo{pages}{107902} (\bibinfo{year}{2004}).

\bibitem[{\citenamefont{Benatti et~al.}(2012)\citenamefont{Benatti, Floreanini,
  and Marzolino}}]{Benatti2012}
\bibinfo{author}{\bibfnamefont{F.}~\bibnamefont{Benatti}},
  \bibinfo{author}{\bibfnamefont{R.}~\bibnamefont{Floreanini}},
  \bibnamefont{and}
  \bibinfo{author}{\bibfnamefont{U.}~\bibnamefont{Marzolino}},
  \bibinfo{journal}{Phys. Rev. A} \textbf{\bibinfo{volume}{85}},
  \bibinfo{pages}{042329} (\bibinfo{year}{2012}).

\bibitem[{\citenamefont{Benatti
  et~al.}(2014{\natexlab{b}})\citenamefont{Benatti, Floreanini, and
  Marzolino}}]{Benatti2014b}
\bibinfo{author}{\bibfnamefont{F.}~\bibnamefont{Benatti}},
  \bibinfo{author}{\bibfnamefont{R.}~\bibnamefont{Floreanini}},
  \bibnamefont{and}
  \bibinfo{author}{\bibfnamefont{U.}~\bibnamefont{Marzolino}},
  \bibinfo{journal}{Phys. Rev. A} \textbf{\bibinfo{volume}{89}},
  \bibinfo{pages}{032326} (\bibinfo{year}{2014}{\natexlab{b}}).

\bibitem[{\citenamefont{Mazza et~al.}(2015)\citenamefont{Mazza, Rossini, Fazio,
  and Endres}}]{Mazza_1367-2630-17-1-013015}
\bibinfo{author}{\bibfnamefont{L.}~\bibnamefont{Mazza}},
  \bibinfo{author}{\bibfnamefont{D.}~\bibnamefont{Rossini}},
  \bibinfo{author}{\bibfnamefont{R.}~\bibnamefont{Fazio}}, \bibnamefont{and}
  \bibinfo{author}{\bibfnamefont{M.}~\bibnamefont{Endres}},
  \bibinfo{journal}{New J. Phys.} \textbf{\bibinfo{volume}{17}},
  \bibinfo{pages}{013015} (\bibinfo{year}{2015}).

\bibitem[{\citenamefont{Iemini et~al.}(2015)\citenamefont{Iemini, Maciel, and
  Vianna}}]{Iemini_PhysRevB.92.075423}
\bibinfo{author}{\bibfnamefont{F.}~\bibnamefont{Iemini}},
  \bibinfo{author}{\bibfnamefont{T.~O.} \bibnamefont{Maciel}},
  \bibnamefont{and} \bibinfo{author}{\bibfnamefont{R.~O.}
  \bibnamefont{Vianna}}, \bibinfo{journal}{Phys. Rev. B}
  \textbf{\bibinfo{volume}{92}}, \bibinfo{pages}{075423}
  (\bibinfo{year}{2015}).

\bibitem[{\citenamefont{Buscemi and Bordone}(2011)}]{Buscemi2011_tripbosfer}
\bibinfo{author}{\bibfnamefont{F.}~\bibnamefont{Buscemi}} \bibnamefont{and}
  \bibinfo{author}{\bibfnamefont{P.}~\bibnamefont{Bordone}},
  \bibinfo{journal}{Phys. Rev. A} \textbf{\bibinfo{volume}{84}},
  \bibinfo{pages}{022303} (\bibinfo{year}{2011}).

\bibitem[{\citenamefont{de~Falco and Tamascelli}(2006)}]{dt06}
\bibinfo{author}{\bibfnamefont{D.}~\bibnamefont{de~Falco}} \bibnamefont{and}
  \bibinfo{author}{\bibfnamefont{D.}~\bibnamefont{Tamascelli}},
  \bibinfo{journal}{J. Phys. A} \textbf{\bibinfo{volume}{39}},
  \bibinfo{pages}{5873} (\bibinfo{year}{2006}).

\bibitem[{\citenamefont{Scott et~al.}(1994)\citenamefont{Scott, Eilbeck, and
  Gilh{\o}j}}]{scott1994quantum}
\bibinfo{author}{\bibfnamefont{A.}~\bibnamefont{Scott}},
  \bibinfo{author}{\bibfnamefont{J.}~\bibnamefont{Eilbeck}}, \bibnamefont{and}
  \bibinfo{author}{\bibfnamefont{H.}~\bibnamefont{Gilh{\o}j}},
  \bibinfo{journal}{Physica D} \textbf{\bibinfo{volume}{78}},
  \bibinfo{pages}{194} (\bibinfo{year}{1994}).

\bibitem[{\citenamefont{Valiente}(2010)}]{Valiente2010twobody}
\bibinfo{author}{\bibfnamefont{M.}~\bibnamefont{Valiente}},
  \bibinfo{journal}{Phys. Rev. A} \textbf{\bibinfo{volume}{81}},
  \bibinfo{pages}{042102} (\bibinfo{year}{2010}).

\bibitem[{\citenamefont{Girolami and Adesso}(2011)}]{Girolami2011negat}
\bibinfo{author}{\bibfnamefont{D.}~\bibnamefont{Girolami}} \bibnamefont{and}
  \bibinfo{author}{\bibfnamefont{G.}~\bibnamefont{Adesso}},
  \bibinfo{journal}{Phys. Rev. A} \textbf{\bibinfo{volume}{84}},
  \bibinfo{pages}{052110} (\bibinfo{year}{2011}).

\bibitem[{\citenamefont{Lee et~al.}(2003)\citenamefont{Lee, Chi, Oh, and
  Kim}}]{Lee2003negat}
\bibinfo{author}{\bibfnamefont{S.}~\bibnamefont{Lee}},
  \bibinfo{author}{\bibfnamefont{D.~P.} \bibnamefont{Chi}},
  \bibinfo{author}{\bibfnamefont{S.~D.} \bibnamefont{Oh}}, \bibnamefont{and}
  \bibinfo{author}{\bibfnamefont{J.}~\bibnamefont{Kim}},
  \bibinfo{journal}{Phys. Rev. A} \textbf{\bibinfo{volume}{68}},
  \bibinfo{pages}{062304} (\bibinfo{year}{2003}).

\bibitem[{\citenamefont{Piccinini et~al.}(2017)\citenamefont{Piccinini,
  Benedetti, Siloi, Paris, and Bordone}}]{gpu17}
\bibinfo{author}{\bibfnamefont{E.}~\bibnamefont{Piccinini}},
  \bibinfo{author}{\bibfnamefont{C.}~\bibnamefont{Benedetti}},
  \bibinfo{author}{\bibfnamefont{I.}~\bibnamefont{Siloi}},
  \bibinfo{author}{\bibfnamefont{M.~G.~A.} \bibnamefont{Paris}},
  \bibnamefont{and} \bibinfo{author}{\bibfnamefont{P.}~\bibnamefont{Bordone}},
  \bibinfo{journal}{Comp. Phys. Comm.} \textbf{\bibinfo{volume}{215}},
  \bibinfo{pages}{235 } (\bibinfo{year}{2017}).

\bibitem[{\citenamefont{Chattaraj and Krems}(2016)}]{chattaraj2016long}
\bibinfo{author}{\bibfnamefont{T.}~\bibnamefont{Chattaraj}} \bibnamefont{and}
  \bibinfo{author}{\bibfnamefont{R.~V.} \bibnamefont{Krems}},
  \bibinfo{journal}{Phys. Rev. A} \textbf{\bibinfo{volume}{94}},
  \bibinfo{pages}{023601} (\bibinfo{year}{2016}).

\end{thebibliography}
\end{document}